\documentclass[reprint,amsmath,amssymb,superscriptaddress,aps,prd,floatfix,onecolumn]{revtex4-1} 

\usepackage{graphicx}
\usepackage{float}
\usepackage{physics}
\usepackage[dvipsnames]{xcolor}
\usepackage[shortlabels]{enumitem}
\usepackage{bigints}
\usepackage{multirow}

\usepackage[normalem]{ulem}

\newcommand{\lat}[2]{{{#1}^3\!\!\times\!\!{#2}}}

\DeclareMathOperator{\asinh}{arsinh}
\allowdisplaybreaks
\begin{document}



\title{Finite density lattice QCD without extrapolation:\\Bulk thermodynamics with physical quark masses from the canonical ensemble}

\author{Alexander Adam}
\affiliation{Department of Physics, Wuppertal University, Gaussstr.  20, D-42119, Wuppertal, Germany}

\author{Szabolcs Bors\'anyi}
\affiliation{Department of Physics, Wuppertal University, Gaussstr.  20, D-42119, Wuppertal, Germany}

\author{Zolt\'an Fodor}
\affiliation{Pennsylvania State University, Department of Physics, State College, PA 16801, USA}
\affiliation{Pennsylvania State University, Institute for Computational and Data Sciences, University Park, PA 16802, USA}
\affiliation{Department of Physics, Wuppertal University, Gaussstr.  20, D-42119, Wuppertal, Germany}
\affiliation{Institute  for Theoretical Physics, ELTE E\"otv\"os Lor\' and University, P\'azm\'any P. s\'et\'any 1/A, H-1117 Budapest, Hungary}
\affiliation{J\"ulich Supercomputing Centre, Forschungszentrum J\"ulich, D-52425 J\"ulich, Germany}

\author{Jana N. Guenther}
\affiliation{Department of Physics, Wuppertal University, Gaussstr.  20, D-42119, Wuppertal, Germany}

%
\author{Ludovica Pirelli}
\affiliation{Department of Physics, Wuppertal University, Gaussstr.  20, D-42119, Wuppertal, Germany}

\author{Paolo Parotto}
\affiliation{Dipartimento di Fisica, Universit\`a di Torino and INFN Torino, Via P. Giuria 1, I-10125 Torino, Italy}

\author{Attila P\'asztor}
\affiliation{Institute  for Theoretical Physics, ELTE E\"otv\"os Lor\' and University, P\'azm\'any P. s\'et\'any 1/A, H-1117 Budapest, Hungary}
\affiliation{MTA-ELTE Lendület "Momentum" Strongly Interacting Matter Research Group, Budapest, Hungary}

\author{Chik Him Wong}
\affiliation{Department of Physics, Wuppertal University, Gaussstr.  20, D-42119, Wuppertal, Germany}

\date{\today}


\begin{abstract}
Quantum Chromodynamics (QCD) at finite density is most often
formulated on the lattice as a grand canonical ensemble. 
Since lattice QCD has a complex action problem at finite baryo-chemical
potential ($\mu_B$), its results at finite density are indirect: e.g.
in the form of a set of expansion coefficients.
In contrast, the canonical formulation offers direct results
for integer-valued net-baryon number. In this work we present
for the first time results in the canonical formulation
with physical quark masses. To this end we use a high statistics
finite-volume lattice ($16^3\times8$) data set that we generated at $\mu_B=0$
with our 4HEX staggered action.
We extend the canonical ensemble to non-integer net-baryon number
and connect the results back to the grand canonical ensemble. 
Unlike reweighing to real $\mu_B$, this method can also be
used with rooted staggered quarks. 
For densities where the sign problem can be overcome by brute force
computing power, this scheme provides lattice QCD results (e.g. for pressure, baryon density) directly, without relying on
any extrapolation in the baryo-chemical potential. In this work we chart the phase diagram by studying bulk thermodynamic observables, which we show to be feasible up to $\mu_B\approx500$~MeV.
\end{abstract}

\maketitle

\section{Introduction\label{sec:intro}} 

Through the gravitational wave observations we gained access
to the densest and hottest matter that our Universe presents,
merging neutron stars. Laboratory experiments of colliding
nuclei interpolate between these conditions and those in the
much hotter Early universe with almost vanishing net-baryon density.
Most notably, the Large Hadron Collider (LHC) at CERN and
the Relativistic Heavy Ion Collider (RHIC) at BNL, have recently explored a
broad range of the phase boundary between the high temperature quark gluon
plasma and the colder hadronic phase. The thermodynamical parameters
of the cooling plasma at the instant of the last inelastic scattering,
the chemical freeze-out, was argued to give accurate estimates on the phase
boundary itself, at least for high collision energies
\cite{BraunMunzinger:2003zz}.

Theoretical solutions to the underlying theory, quantum chromodynamics (QCD),
later confirmed the good agreement between the chemical freeze-out and
the chiral crossover. Most prominently, lattice QCD was able to establish
the fact that the transition is a crossover \cite{Aoki:2006we}, its temperature
at zero density \cite{Aoki:2006br,Aoki:2009sc,Borsanyi:2010bp,Bazavov:2011nk}, 
and the crossover line on the plane of 
temperature ($T$) and baryo-chemical potential ($\mu_B$) \cite{Bonati:2015bha,Bellwied:2015rza,Bazavov:2018mes,Borsanyi:2020fev}.
Other theoretical tools, too, have been successfully employed to
compute the cross-over line, e.g. in the Dyson-Schwinger formalism
\cite{Fischer:2018sdj,Gao:2020fbl,Gunkel:2021oya} or with the functional
renormalization group \cite{Fu:2019hdw}.

The interest in hot and dense QCD matter is further fuelled by
the prospect of the discovery of a critical point on the chiral cross-over
line \cite{Stephanov:1998dy}. Several effective models
support the existence of such a critical point,
e.g. random matrix, quark-meson model or Nambu-Jona-Lasino, inculding
their extensions with the Polyakov loop. Quite interestingly,
first principles functional methods, too, give predictions to the location
of this conjectured point \cite{Gao:2020fbl,Gunkel:2021oya,Fu:2019hdw},
all in the ballpark of $\mu_B\approx 650$~MeV, $T\approx 100$~MeV.
See Ref.~\cite{Fischer:2026uni} for a recent summary.

Lattice QCD can access this range only indirectly. In the standard
formulation, the quark degrees of freedom are integrated out to form
the so called quark determinant.
The latter depends on the quark mass, the chemical potential and the background
gauge field, which is then sampled with the weight given by the gauge 
action and this determinant. Unfortunately, unless the chemical potential is
either zero or purely imaginary, the determinant becomes a complex number,
unsuitable for a probability to be used in importance sampling. Perhaps the
most successful indirect methods are the Taylor expansion, where derivatives
of the grand potential and other observables are computed at $\mu_B=0$
\cite{Gavai:1989ce,Allton:2002zi}, and the analytical continuation from
imaginary chemical potential
\cite{Alford:1998sd,deForcrand:2002ci,DElia:2002gd}.  Both methods depend on
favorable analytical features of the grand potential.  Physics results at
finite density are obtained in form of truncated extrapolation formulae with lattice data as input. By today continuum extrapolated results are available up to
second \cite{Borsanyi:2011sw,Bazavov:2012jq,Bollweg:2021vqf}, fourth
\cite{Bellwied:2015lba,Ding:2015fca}, sixth
\cite{Borsanyi:2023wno,Bollweg:2022fqq} order of the $\mu_B$ expansion and even
eighth \cite{Borsanyi:2018grb,Bollweg:2022fqq,Bollweg:2022rps} and tenth orders
\cite{Adam:2025hpb} at finite lattice spacing. While the Taylor coefficients
are computed at $\mu_B=0$, thus, free from the complex action problem, they are
hit by a cancellation (or sign) problem, with exponentially increasing severity with volume and
order \cite{Adam:2025hpb}.

The idea of reweighting from $\mu_B=0$ to the physical $\mu_B>0$ parameter
space was a natural suggestion \cite{Barbour:1997ej}. This, however, introduces
a sign problem and an overlap problem. The latter could be mitigated using
multi-parameter reweighting \cite{Fodor:2001au,Fodor:2001pe}
or a density-of-states approach \cite{Fodor:2007vv}. In a more
advanced setup with finer lattices the overlap problem was surprisingly mild in
small volumes \cite{Borsanyi:2022soo}.  Let us point out, that the Taylor
coefficients, when computed from a single $\mu_B=0$ ensemble, are equivalent to
those of the reweighted ensemble. In this viewpoint, the Taylor expanded finite
density results are mere approximations of what a full reweighting would give,
and inherit a sign problem and an overlap problem of the same magnitude.

The sign problem is only one aspect of the difficulties with reweighting. Even if the
additional overlap problem is treated e.g. by employing sign quenched simulations
\cite{Giordano:2020roi,Borsanyi:2021hbk,Borsanyi:2022soo}, and we own
sufficient computer resources to control the sign problem, there is a further
obstacle. The overwhelming majority of lattice
thermodynamics results is simulated in the staggered formulation, especially
so for physical quark masses. The rooting procedure at finite chemical
potential is ambiguous, leading to severe lattice artefacts
\cite{Golterman:2006rw} including a fake transition at
$\mu_B\approx m_\pi 3/2$ \cite{Barbour:1997bh,Borsanyi:2023tdp}.
This paper offers a solution to the rooting problem through a detour to the
canonical formalism.

The canonical ensemble does not easily lend itself for simulation on the lattice.
While it is well known how to introduce a baryo-chemical potential such that it
respects the underlying $U(1)_V$ symmetry \cite{Hasenfratz:1983ba},
one cannot associate a baryon number ($N$) to a given gauge configuration, once
the quark degrees of freedom are integrated out. Each configuration contributes
to every baryon number $N$ with some complex weight \cite{Alexandru:2014hga}. Specifically
the $N=0$ sector can and has been simulated without a sign problem
\cite{Kratochvila:2006jx}.  The $N>0$ ensembles can be obtained e.g. by 
reweighting from the grand canonical $\mu_B=0$ ensemble
\cite{deForcrand:2006ec}. Importance sampling at finite density was introduced
in Ref.~\cite{Alexandru:2005ix} in a sign-quenched setup.
Early studies demonstrate that the easily accessible canonical
observable $\mu_B(N)$ marks the first order finite density transition
with an unambiguous 'S'-shape pattern \cite{deForcrand:2006ec,Li:2010qf}
from which a strategy to locate the critical endpoint can be derived
\cite{Li:2011ee}.

Since the severity of the finite density sign problem is strongly linked to the pion mass
\cite{Alexandru:2014hga}, reaching the physical point, or any realistic lattice
spacing has not been feasible, previously.  Yet, many technical issues could be
discussed in detail, e.g. the deterministic
\cite{Hasenfratz:1991ax,deForcrand:2006ec} computation of the complex weight
associated with $N$ baryon numbers, the so called canonical determinant,
or its numerically stable construction through a winding number expansion
\cite{Li:2010qf,Fukuda:2015mva}. 

A further challenge posed by fixed-$N$ canonical data is the extraction of
grand canonical information, such as the fluctuations of conserved charges 
\cite{Danzer:2012vw}. At finite $\mu_B$ each $Z_N$ sector contributes
with the weight $e^{N\mu_B}$. Thus, grand canonical observables, such as 
moments of $N$ are the results of an infinite sum in $N$.  
The inevitable truncation in $N$ leads to systematic errors especially for
higher moments of $N$. This is unfortunate near the physical point, where one
quickly loses control for higher $N$ sectors.

The aim of this work is twofold. With simulations at the physical point
on a finite lattice we demonstrate the practical feasibility of the
canonical approach. We will work with a physical system size of $LT=2$.
In this volume (ca.~20~fm$^3$ at $T=145~\textrm{MeV}$) the sign problem can be
addressed with high statistics ensembles up to $\mu_B\lesssim 500$~MeV
in the crossover range. We can relate the phenomenological findings
to the expected Skellam distribution of a non-interacting hadron gas.
We briefly discuss how the strangeness number can be added to the formalism.

The second aim of this work is to give a somewhat technical discussion
on how the partition function is obtained from a single $\mu_B=0$ ensemble,
including all center sectors, and how the staggered rooting problem is
circumvented at the same time. 
The method to recover grand canonical results from canonical lattice
data without performing a truncated sum in $N$ is a 
further novelty in this work.
The final results on the QCD
pressure and chemical potential as a function of the baryon density are
analogous to the would-be outcome of reweighting, as if the complex rooting
problem of Refs.~\cite{Golterman:2006rw,Borsanyi:2023tdp} did not exist.

The exciting feature of these results is the absence of an extrapolation in $\mu_B$.
The phenomenological results and theoretical arguments of this work then suggest
an optimistic prospect: lattice QCD is no longer a mere tool to provide
extrapolation coefficients, but it can also give actual truncation-free
finite density results. The sign problem itself can be dealt with, thanks to today's
computer power, up to a certain density. 
Thus, finite density simulations are indeed possible, even
with physical parameters and with a lattice resolution $(N_\tau=8)$ that is
suitable for phenomenological studies, covering much of the parameter space accessible to the RHIC Beam Energy Scan program.

\begin{figure}
\centering
\includegraphics[width=0.48\textwidth]{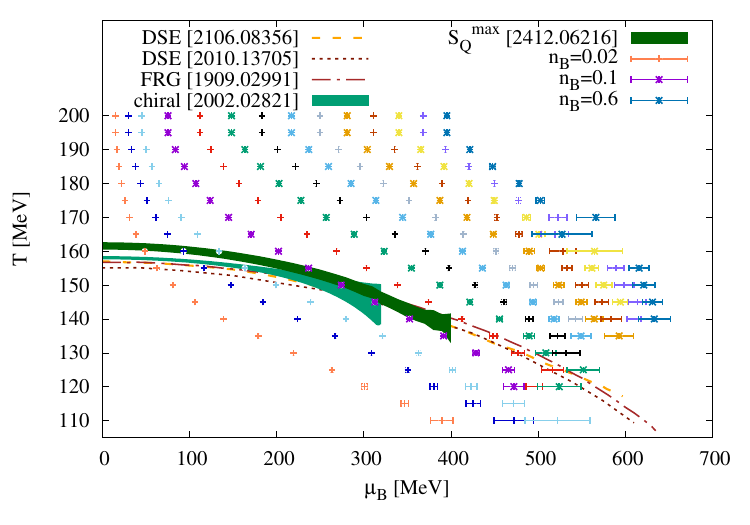}
\caption{\label{fig:can_phase_diag_inf_alpha}
Contours of constant net-baryon density on the QCD phase diagram.
Each color represents a fixed density, from left to right $n_B=$
0.02, 0.04, 0.06, 0.10, 0.15, 0.20, 0.25, 0.30, 0.35, 0.40, 0.45, 0.50, 0.55, $0.60~\mathrm{fm}^{-3}$.
The errors are statistical.
}
\end{figure}

In Fig.~\ref{fig:can_phase_diag_inf_alpha} we give an example on what is possible
with our methodological improvements.  We show the results of our
high statistics lattice simulations as contours of fixed baryon density
on the $T-\mu_B$ plane of the QCD phase diagram. At present, we can follow these
up to a chemical potential of ca. 500 MeV. At some higher chemical potential
these lines will cross, if the transition becomes first order. The knowledge
of these lines also implies that the equation of state is known at finite density.


This paper starts with revisiting the canonical formalism in section
\ref{sec:canon}. The basic concepts are illustrated with the example of a
non-interacting hadron gas (see appendix \ref{app:skellam}).
In section~\ref{sec:algo} we propose new algorithmic ingredients that
optimize the numerical procedure. The first 
phenomenological results in the canonical formalism are presented in section
\ref{sec:thermo}, mostly focusing on the pressure and chemical potential as
observables. Here, the direct results at fixed (canonical) baryon number are 
specific to the system volume. We generalize the discussion to an arbitrary
volume and connect back to the grand canonical ensemble in section
\ref{sec:grand}. We conclude with a discussion on future directions in section \ref{sec:prospects}.

\section{The canonical formalism \label{sec:canon}} 

\subsection{Generic thermodynamics}

At the energy scale of the QCD transition the theory features three exactly conserved
charges, which can be seen as the net quark number for each of the up, down and strange flavors, or equivalently the baryon number B, electric charge Q and strangeness S.
For the purpose of this paper we do not consider heavier quarks, hence our dynamical
lattice simulations will be performed with $N_f=2+1$ flavors, and we introduce the quark chemical potentials in the
standard exponential form~\cite{Hasenfratz:1983ba}. We focus on two of said conserved charges, namely the net baryon number $N_B$ and the net strangeness
$N_S$. 

In textbooks on can find a simple relation between the canonical ($Z_C$) and grand
canonical partition functions ($Z_{GC}$):
\begin{eqnarray}
Z_{GC}(T,V,\mu) = \sum_{N=-\infty}^{\infty} e^{\frac{\mu}{T} N} Z_C(T,V,N) \, \, .
\label{eq:ZGCrel}
\end{eqnarray}
For simplicity, here and in the following discussions we use generic $\mu$ and $N$ as
conjugate variables, standing for chemical potential and net particle number.
Note that Eq.~(\ref{eq:ZGCrel}) features a summation over 
integer-valued $N$, as in fact
there is no state in any finite box with a fractional baryon number (or
strangeness). This remains true in lattice QCD, also with rooted staggered
quarks, since this feature is protected by the unbroken center symmetry $\mathbb{Z}_3$ of the gauge group $SU(3)$. A consequence of this symmetry
is the periodicity in the imaginary-valued chemical potential $\mu/T=i\varphi$, which is evident from the relation
\begin{eqnarray}
Z_{GC}(T,V,i\varphi T) = \sum_{N=-\infty}^{\infty} e^{i\varphi N} Z_C(T,V,N) \, \, .
\label{eq:ZGCrelimmu}
\end{eqnarray}
A more practical, equivalent form of Eq.~(\ref{eq:ZGCrelimmu}) can be obtained by
inverting the Fourier series as
\begin{eqnarray}
Z_C(T,V,N)&=&  \int_0^{2\pi} \frac{d\varphi}{2\pi} e^{-i\varphi N} Z_{GC}(T,V,i\varphi T)\,,
\label{eq:ZCfourier}
\end{eqnarray}
or in terms of the grand potential ($\Omega$)
 and the Helmholtz free energy ($F$)
\begin{eqnarray}
\Omega(T,V,\mu)&=&-T\log Z_{GC}(T,V,\mu)\,,\\
F(T,V,N)&=&-T\log Z_C(T,V,N)\,,
\label{eq:FfourierA}
\end{eqnarray}
which inserting Eq.~\eqref{eq:ZGCrelimmu} yields:
\begin{eqnarray}
    F(T,V,N)&=& -T \log \int_0^{2\pi} \frac{d\varphi}{2\pi} e^{-i\varphi N - \frac{1}{T}\Omega(T,V,i\varphi T)} \, \, .
\label{eq:FfourierB}
\end{eqnarray}
The right hand side of Eq.~(\ref{eq:FfourierB}) can be directly implemented in lattice QCD, since only
imaginary-valued chemical potentials are used.
Note that the canonical free energy ($F$) of Eq.~(\ref{eq:FfourierB}) is defined only at
integer values of $N$.

Another route from $\Omega(T,V,\mu)$ to $F(T,V,N)$
employs the Legendre transform
\begin{eqnarray}
F_{LT}(T,V,\langle N\rangle)=\Omega(T,V,\mu) + \mu \langle N\rangle\,, \qquad \qquad \rm{with} \quad \langle N\rangle = -\frac{\partial \Omega}{\partial \mu}\,
\end{eqnarray}
and a function $F_{LT}$ ($LT$ stands for Legendre transform) depending on the unconstrained 
particle number $\langle N\rangle$,
expected to agree with $F(T,V,N)$ only in the thermodynamic limit.

For the purpose of the paper it is instructive to derive the relation
between $F(T,V,N)$ and $F_{LT}(T,V,\langle N\rangle)$. For this we
will suppress some of the function arguments.
Let us imagine that the system described by $\Omega(\mu)$ is
replicated $\alpha$ times, thereby rescaling 
$\Omega$, $V$ and $N$:
\begin{eqnarray}
F(T,\alpha V,\alpha N)&=& -T \log 
\int_0^{2\pi} \frac{d\varphi}{2\pi} e^{-i\varphi \alpha N  - \frac{\alpha}{T}\Omega(i\varphi T)}\,
\label{eq:FLTderiv1}
\end{eqnarray}
where $\alpha N$ is an integer.
Because we do not require the replica factor $\alpha$ to be an integer,
$N$ in Eq.~(\ref{eq:FLTderiv1}) can also be fractional. Then,
$F/\alpha$ will stand for the canonical free energy averaged over the
replicas, which together account for $\alpha N$ particles.  
In the following derivation we demonstrate that the Legendre transformed
free energy can be obtained in the following limit:
\begin{eqnarray}
F_{LT}(T,V,N)&=& \frac{1}{\alpha}F(T,\alpha V,\alpha N) 
+\frac{T}{2}\alpha^{-1}\log{\alpha^{-1}}
+\mathcal{O}\left(\frac{1}{\alpha}\right)
\label{eq:FLTasymp2}
\end{eqnarray}

Let us denote with $\mu^\star$ 
the chemical potential solving the equation
\begin{eqnarray}
N = -\left.\frac{\partial \Omega}{\partial\mu}\right|_{\mu=\mu^\star}\,.
\end{eqnarray}
In this case, at $i\varphi^\star=\mu^\star/T$ the $\varphi$-derivative of the exponent in Eq.~(\ref{eq:FLTderiv1}) is zero: in other words, $\varphi^\star$ defines a saddle point.
Note that, although for real $N$ $\varphi^\star$ will be imaginary, we consider a generic complex $\varphi^\star$.
First adjusting the boundaries of the periodic integral, then
expanding around the saddle point $\varphi^\star$ one finds
\begin{eqnarray}
\frac{1}{\alpha}F(\alpha N)&=&-\frac{T}{\alpha}
\log\int_{\mathrm{Re}\varphi^\star-\pi}^{\mathrm{Re}\varphi^\star+\pi} 
\frac{d\varphi}{2\pi}
e^{-i\varphi^\star\alpha N - \frac{\alpha}{T} \Omega(i\varphi^\star T) }\nonumber\\
&&\quad\quad\quad\cdot
e^{-\alpha VT^3 \sum_{n=2}^{\infty}\frac{\chi_n(\varphi^\star)}{n!} (\varphi-\varphi^\star)^n} \, \, ,
\label{eq:FLTderiv2}
\end{eqnarray}
where we introduced the standard notation for the generalized (baryon) susceptibilities:
\begin{eqnarray}
\chi_n &=& \frac{1}{VT^3} \frac{\partial^n\log Z_{GC}}{\partial (\mu/T)^n}\,.
\end{eqnarray}

Now, in the first exponent of Eq.~(\ref{eq:FLTderiv2}) we recognize $F_{LT}(N)$, which is independent of the integrand and can be pulled out as
\begin{eqnarray}
\frac{1}{\alpha}
F(\alpha N)&=&F_{LT}(N) 
-\frac{T}{\alpha} 
\log \frac{1}{2\pi\sqrt{\alpha}} 
 \int_{-\pi\sqrt{\alpha}}^{\pi\sqrt{\alpha}}  dx \exp\biggl[\nonumber\\
&&
- VT^3 \sum_{n=2}^{\infty}\frac{\chi_n(\varphi^\star)}{n!\alpha^{n/2-1}} (x - \sqrt{\alpha} i\mathrm{Im}\varphi^\star)^n
\biggr]
\label{eq:FLTderiv3}
\end{eqnarray}
where we substituted $x=\sqrt{\alpha}(\varphi-\mathrm{Re}\varphi^\star)$.
Since this definite Gaussian integral converges to 
$\int^{\infty}_{-\infty}dx$ faster than any power as $\alpha\to\infty$, being interested in the leading asymptotic behavior we can safely
extend the integral limits. This then allows us to shift the integration contour via
$z=x-i\sqrt{\alpha}\mathrm{Im}\varphi^\star$.
At this point the leading Gaussian integral asymptotically gives $\sqrt{\frac{2\pi}{VT^3 \chi_2(\varphi^\star)}}$. 
while the sub-leading terms can be further expanded in 
$1/\sqrt{\alpha}$ as:
\begin{eqnarray}
\int_{-\infty}^{\infty} dz
e^{-\frac{VT^3\chi_2}{2} z^2}
\prod_{n=3}^{\infty}
\sum_{k=0}^\infty\frac{1}{k!}
\left(\frac{z}{\sqrt{\alpha}}\right)^{nk}
\left[\frac{-VT^3\alpha\chi_n}{n!}\right]^k \, \, .
\nonumber
\end{eqnarray}

Here, odd terms in $\sqrt{\alpha}$ will evaluate to zero, hence the
leading correction will be of order $1/\alpha$ (one term is
$\sim(\chi_3)^2$ and one is $\sim\chi_4$).

In conclusion, we can express the Legendre-transformed $F_{LT}$ as a limit with
known asymptotic behavior:
\begin{eqnarray}
\frac{1}{\alpha}F(T,\alpha V,\alpha N)&=& F_{LT}(T,V,N)
+\frac{T}{2\alpha}\log\left(2\pi\alpha\chi_2(\mu^\star)VT^3\right)\left(1+\mathcal{O}(\alpha^{-1})\right)\,,
\label{eq:FLTasymp}
\end{eqnarray}
as announced in Eq.~(\ref{eq:FLTasymp2}).
The $\alpha^{-1}\log{\alpha^{-1}}$ term also converges to zero in the infinite $\alpha$ limit, and drops out from the
$V$ and $N$ derivatives. In section \ref{sec:grand} we will use
Eq.~(\ref{eq:FLTasymp2}) with several $\alpha$ values to numerically perform the
$\alpha\to\infty$ extrapolation and recover $F$.

In this paper we will focus on the charting of the QCD phase 
diagram, and in particular on bulk observables like the 
chemical potential $\mu$, the pressure and its derivatives 
with respect to $\mu$ itself. The chemical
potential is easily found in the Legendre transform formalism
via
\begin{eqnarray}
\mu(T,V,N) = \frac{\partial F_{LT}(T,V,N)}{\partial N} \, \, .
\end{eqnarray}
In the case of $F(T,V,N)$ with integer-valued $N$ this is usually understood
as a discrete derivative of the form
\begin{eqnarray}
\mu^f&=&F(T,V,N+1)-F(T,V,N) \label{eq:muf}\,,\\
\mu^b&=&F(T,V,N)-F(T,V,N-1) \label{eq:mub} \, ,
\end{eqnarray}
where the indices $f$ and $b$ stand for forward and backward derivatives.  Some
authors in the lattice community prefer $\mu^f$
\cite{Alexandru:2005ix,Li:2011ee}, some use $\mu^b$ \cite{deForcrand:2006ec}, but in a sufficiently large volume ($V\to\infty$ with constant density $N/V$)
both derivatives approach $\mu(T,V,N)$ as $1/V$, as one can easily see from Eq.~(\ref{eq:FLTasymp}). This is under the assumption that
the volume asymptotics in the grand canonical $\Omega(T,V,\mu)$ is faster than
$1/\alpha$. In fact, in the absence of a massless degree of freedom,
one expects an exponential approach to the thermodynamic limit.

To define the grand canonical pressure it is common to make the assumption of perfect homogeneity, i.e.:
\begin{eqnarray}
p_{GC} = -  \frac{\partial \Omega}{\partial V} \approx - \frac{\Omega}{V}
= \frac{T}{V}\log Z_{GC}\,,
\label{eq:GChom}
\end{eqnarray}
and the volume derivative for the pressure in the canonical 
setup can only be performed under the same assumption, that is:
\begin{eqnarray}
\Omega(T,\alpha V,\mu) &=& \alpha \Omega(T,V,\mu)\,,\\
F(T,\alpha V, N)&=&-T\log\int_0^{2\pi} 
\frac{d\varphi}{2\pi}
e^{-i\varphi N - \alpha T^{-1} \Omega(T,V,i\varphi T)}\,.
\end{eqnarray}
With this volume dependence the canonical pressure reads
\begin{eqnarray}
p_C(T,V,N)&=&
-\frac{\partial F(T,V,N)}{\partial V}
=-\frac{1}{V}\left.\frac{d F(T,\alpha V,N)}{d\alpha}\right|_{\alpha=1}~~\nonumber\\
&=&-\frac{T}{V}
\frac{
\int_0^{2\pi} \frac{d\varphi}{2\pi}
\Omega(i\varphi T) e^{-i\varphi N - T^{-1}\Omega(T,V,i\varphi T)}
}{
\int_0^{2\pi} \frac{d\varphi}{2\pi}
 e^{-i\varphi N -T^{-1} \Omega(T,V,i\varphi T)}
} \, \, .
\label{eq:pcanon}
\end{eqnarray}

So far we have not fixed the normalization of $Z_{GC}$ and $Z_{C}$.
What we will actually compute in section \ref{sec:algo} is the ratio
\begin{eqnarray}
Z_{GCR}(\varphi)= Z_{GC}(\mu=i\varphi T)/Z_{GC}(\mu=0)
\label{eq:ZGCrat}
\end{eqnarray}
(GCR stands for grand canonical ratio). Then, Eq.~(\ref{eq:FfourierB}) will give us 
$Z_{C}(N)/Z_{GC}(\mu=0)$ as the Fourier transform of Eq.~(\ref{eq:ZGCrat}). 
This means that this formalism can readily provide the QCD pressure relative to the grand canonical
$\mu=0$ result, i.e.:
\begin{eqnarray}
&&\Delta p_C(T,V,N) = p_C(T,V,N)- p_{GC}(T,V,\mu=0)\nonumber\\
&&
=\frac{T}{V}
\frac{
\int_0^{2\pi} \frac{d\varphi}{2\pi}
Z_{GCR}(\varphi) \log Z_{GCR}(\varphi) e^{-i\varphi N}
}{
\int_0^{2\pi} \frac{d\varphi}{2\pi}
Z_{GCR}(\varphi)
 e^{-i\varphi N }
}
\label{eq:dpcanon}
\end{eqnarray}

This last equation is convenient for lattice implementation, and we will show our corresponding results in section \ref{sec:thermo}. Note that at $N=0$ this is
negative and proportional to $\sim 1/V$.

Another quantity of great interest in the study of the QCD phase diagram is the fluctuation of $N$ (the net baryon number), which is a grand canonical
observable ($\chi_2$). Yet, it is possible to extract this from $F_{LT}$ as
\begin{eqnarray}
\frac{\partial^2}{\partial N^2} F_{LT}(T,V,N)&=& \frac{T}{(VT^3)\chi_2}\,,
\end{eqnarray}
or, starting from $F$ and with integer $N$:
\begin{eqnarray}
\frac{1}{\chi_2} &=& VT^2 \left(F(T,V,N+1)-2F(T,V,N)+F(T,V,N-1)\right) \nonumber\\
&&+ 
\mathcal{O}\left(V^{-1}\right)\,,
\end{eqnarray}
where the asymptotic form is valid at fixed density.

For the interested reader we present a detailed account on the non-interacting
hadron gas in appendix \ref{app:skellam}. There we actually derive the
leading volume effects and show how the thermodynamic limit is reached.
It turns out that in the hadronic phase the finite volume corrections
to the backward derivative approximately cancel, and so it reproduces the grand
canonical $\mu$ already with very small $N$. Of course, the range where this model gives
an adequate description of lattice data is limited to $T<T_c$. Nevertheless, we will use the backward derivative $\mu^b$ of Eq.~(\ref{eq:mub}), motivated by this model,
consistently at all temperatures.


\section{Optimized reweighting to the canonical ensemble\label{sec:algo}} 

We include all technical details on our lattice analysis in this section.
Part \ref{sec:algo:canon} summarizes the main ingredients to go
from the original $\mu_B=0$ ensembles to observables at fixed $N_B$. 
We will also discuss the staggered rooting here.
Part \ref{sec:algo:center} proposes an optimized reweighting scheme
for the correct sampling of the center sectors. Without this step the
errors would be dramatically larger. Readers not interested in these
technical details can safely skip this section.

\subsection{Construction of the canonical determinants\label{sec:algo:canon}} 

The starting point for our study is a set of grand canonical ensembles 
at $\mu_B=0$ on a $16^3\times8$ lattice with physical pion masses, namely one ensemble per temperature in the range $T=110\dots200$~MeV.
Each ensemble was simulated in $\mathcal{O}(100)$ streams with ca. 300k trajectories in every RHMC stream.
The employed 4HEX staggered action and earlier stages of these simulations
have already been documented in Refs.~\cite{Borsanyi:2023wno,Borsanyi:2024xrx}.
The statistics is given in table \ref{tab:stat}.

\begin{table} 
\begin{tabular}{|c|r|p{3ex}p{10ex}p{10ex}p{10ex}|}
\hline
\multirow{2}{4em}{$T$~[MeV]}  & 
\multirow{2}{10ex}{statistics} & \multicolumn{4}{|c|}{center sector distribution [\%]}  \\
&&& 0 & $+2\pi/3$ & $-2\pi/3$ \\
\hline
110 & 410816 && 87.7 &  6.1 &  6.2 \\
115 & 1054391 && 90.8 &  4.6 &  4.6 \\
120 & 1080583 && 93.3 &  3.4 &  3.3 \\
125 & 1512103 && 95.4 &  2.3 &  2.3 \\
130 & 2129599 && 96.9 &  1.5 &  1.5 \\
135 & 1216275 && 98.0 &  1.0 &  1.0 \\
140 & 4002298 && 98.8 &  0.6 &  0.6 \\
145 & 2277535 && 99.3 &  0.35 &  0.34 \\
150 & 1496094 && 99.6 &  0.19 &  0.19 \\
155 & 1622365 && 99.8 &  0.93 &  0.94 \\
160 & 2226025 && 99.9 &  0.045 &  0.044 \\
165 & 1649108 && 100.0 &  0.019 &  0.02 \\
170 & 1166291 && 100.0 &  0.0083 &  0.0086 \\
175 & 1599122 && 100.0 &  0.0027 &  0.0036 \\
180 & 1240882 && 100.0 &  0.0015 &  0.0012 \\
185 & 851014 && 100.0 &  0.00071 &  0.00035 \\
190 & 317898 && 100.0 &  0.00031 &  0.0 \\
195 & 361870 && 100.0 &  0.00028 &  0.0 \\
200 & 323969 && 100.0 &  0.0 &  0.0 \\
\hline
\end{tabular}
\caption{\label{tab:stat}
Statistics of the $\mu_B=0$ ensembles on the $16^3\times 8$ lattice. On each
configuration the full eigenvalue problems of both the strange and light
determinants were solved. We also give the percentage of the gauge
configurations per center sector as defined in section \ref{sec:algo:center}.
}
\end{table} 

There are several approaches in the literature to create a canonical ensemble.
Our approach is most similar to Refs.~\cite{deForcrand:2006ec,Danzer:2012vw}, in
the sense that these works also start with $\mu_B=0$ ensembles.

The grand canonical path integral at imaginary baryo-chemical potential
$\mu_B/T =i \varphi$
in QCD with $N_f$ degenerate flavors reads 
\begin{eqnarray}
Z_{GCR}(\varphi) = \frac{1}{Z_{GC}(\mu=0)}
\int \mathcal{D}U \left(\det M(U,\varphi/3) \right)^{N_f} e^{-\beta S_G(U)}
\end{eqnarray}
where $S_G$ is the gauge action, and $\det~M$ stands for the quark determinant.
Its argument $\varphi/3$ refers to the (imaginary) quark chemical potential.
We work in the staggered formulation with degenerate light flavors and a heavier strange flavor. In this (physical) setup we introduce two phases, $\varphi_B$
and $\varphi_S$ for the baryon and strange chemical potentials, respectively.
\begin{eqnarray}
Z_{GCR}(\varphi_B,\varphi_S) = \frac{1}{Z_{GC}(\mu=0)}
\int \mathcal{D}U 
\left(\det M_{KS}(U,m_{ud}, \varphi_B/3) \right)^{1/2}
\left(\det M_{KS}(U,m_{s}, \varphi_B/3-\varphi_S) \right)^{1/4}
 e^{-\beta S_G(U)}\,.
\label{eq:ZGCR1}
\end{eqnarray}
We wrote the mass arguments ($m_{ud}$ and $m_s$)
of the Kogut-Susskind quark determinants explicitly for clarity.

A crucial feature of Eq.~(\ref{eq:ZGCR1}) is that all determinants
are real and positive for real $\varphi_B$, $\varphi_S$. This means
that no ambiguous rooting is performed on complex numbers. $Z_{GCR}$ will
only ever be evaluated at such real phases. Once the path integral
(or in other words: ensemble average) is taken, we will introduce
real densities in the canonical formalism.

A second important feature of the path integral in Eq.~(\ref{eq:ZGCR1})
is its periodicity both in $\varphi_B$ and in $\varphi_S$:
\begin{eqnarray}
Z_{GCR}(\varphi_B,\varphi_S) &=& Z_{GCR}(\varphi_B,\varphi_S+2\pi)
\label{eq:periodicityS}\\
Z_{GCR}(\varphi_B,\varphi_S) &=& Z_{GCR}(\varphi_B+2\pi,\varphi_S)\,.
\label{eq:periodicityB}
\end{eqnarray}
This was pointed out first in Ref.~\cite{Roberge:1986mm}. Its direct consequence
is the Roberge-Weiss transition at $\mu_B=i\pi T$ at high temperature.
The periodicity is the consequence of two statements:
a) An imaginary quark chemical potential $\varphi_q$ can be interpreted as a
phase factor $e^{i\varphi_q}$ introduced in the periodic 
Euclidean temporal boundary condition for the quark fields. Since
this factor is just 1 for $\varphi_q=2\pi$,
this immediately explains Eq.~(\ref{eq:periodicityS}). 

b) The unbroken center symmetry of the gauge action and the Haar measure 
allows the simultaneous transformation of the gauge fields on a single
time slice by an element of the center group $\mathbb{Z}_3$: $e^{-i2\pi/3}$.
The determinants will maintain their value if all quark
chemical potentials are set forward by $2\pi/3$ at the same time. This
amounts to a change $\varphi_B\to\varphi_B+2\pi$.

Charge conjugation ($\mathcal{C}$) symmetry applies for $\varphi_B$ and $\varphi_S$ together, implying:
\begin{eqnarray}
Z_{GCR}(\varphi_B,\varphi_S) &=& Z_{GCR}(-\varphi_B,-\varphi_S)\,.
\end{eqnarray}
Center symmetry and $\mathcal{C}$-symmetry together shrink the relevant parameter
space to $\varphi_B=0,\dots,\pi$ and $\varphi_S=0,\dots,2\pi$.
This whole range was explored in Ref.~\cite{Bellwied:2021nrt} through a two dimensional
mesh of simulation points in an effort to compute the fugacity
expansion of $Z_{GCR}(\varphi_B,\varphi_S)$. Here, instead, we compute this ratio
by means of reweighting from zero to imaginary chemical potential.
It may sound unnecessary to reweight to a point in the parameter space,
where simulations are easily possible. However,
reweighting has some clear advantages. First, one can evaluate
the determinants at any chemical potential, thus
$Z_{GCR}(\varphi_B,\varphi_S)$ will be available as a continuous function in both arguments.
Second, the statistical correlations between nearby (and
also distant) parameters are fully kept. Without the first feature some
modeling of the data would be necessary, this is precisely what we wish to
avoid in this work. Without the second feature the higher modes in the
Fourier transform of $Z_{GCR}$ would be lost in the noise.

The computation of the quark determinants is a numerical challenge.
We use the reduced matrix formalism for staggered quarks
\cite{Hasenfratz:1991ax}, whereby the quark determinant
are represented by a set of $6 N_x\times N_z\times N_y$ eigenvalues 
$\xi_i$
\begin{eqnarray}
\frac{\det M_{KS}(U,m,\varphi)}{\det M_{KS}(U,m,0)}
= \prod_{i=1}^{6N_s^3}
\frac{\xi_i(m,U) - e^{i\varphi}}{\xi_i(m,U)-1}\,.
\end{eqnarray}
This determinant is manifestly real, thanks to the 
$\gamma_5$-Hermiticity. This implies that every $\xi_i$ eigenvalue has a pair
$\xi_j$ with $\xi_i=1/\xi_j^\star$. This can be also exploited to maintain numerical
stability. The resulting reweighted partition function can then be written
as an ensemble average:
\begin{eqnarray}
Z_{GCR}(\varphi_B,\varphi_S) &=&
\left\langle
\left(\frac{\det M_{KS}(U,m_{ud}, \varphi_B/3)}{\det M_{KS}(U,m_{ud},0)} \right)^{1/2}
\left(\frac{\det M_{KS}(U,m_{s}, \varphi_B/3-\varphi_S) }{\det M_{KS}(U,m_{s},0)}\right)^{1/4}
\right\rangle
\label{eq:ZGCR2}
\end{eqnarray}
The practical usefulness of this formula depends on the knowledge of
the $\xi_i(U,m)$ eigenvalues. Our lattice simulation code computes
the reduced matrix, which is $24576\times24576$ for our $16^3\times8$ staggered
lattices. The 4HEX smearing is applied before the construction of the
matrix, and we do this computation and the subsequent call to
QR-algorithm (and the preparatory transformation to upper Hessenberg form)
separately for the two masses. For the dense linear algebra 
we use the MAGMA linear algebra package \cite{MR1484478}. The main part
of the computer resources used for this project is spent in the MAGMA library.
We carried out these steps for all the configurations listed in
Table \ref{tab:stat}.
We refer to Ref.~\cite{Giordano:2019gev} for further technical details.

Once we know the $2\times 24576$ eigenvalues for each of the
$\mathcal{O}(10^6)$ configurations, the computation of Eq.~(\ref{eq:ZGCR2}) 
is straightforward and requires the computational power of a
modern workstation. We computed $Z_{GCR}(\varphi_B,\varphi_S)$ on 
a grid of $33\times64$ parameters (mapped to $[0,\pi]\times[0,2\pi)$).
Refining the grid did not bring noticeable changes in the result.
Before taking the ensemble averages two manifestly positive
determinant functions were
pre-calculated at $192$ imaginary quark chemicals each, amounting
to 384 real numbers per configuration:
\begin{eqnarray}
D_q(\varphi_q) &=&
\left(\frac{\det M_{KS}(U,m_{ud}, \varphi_q)}{\det M_{KS}(U,m_{ud},0)} \right)^{1/2}
\left(\frac{\det M_{KS}(U,m_{s}, \varphi_q) }{\det M_{KS}(U,m_{s},0)}\right)^{1/4}\,, \label{eq:Dq}\\
D_s(\varphi_s) &=&
\left(\frac{\det M_{KS}(U,m_{s},\varphi_s) }{\det M_{KS}(U,m_{s},0)}\right)^{1/4}\,. \label{eq:Ds}
\end{eqnarray}

The average in Eq.~(\ref{eq:ZGCR2}) is finally transformed to a canonical
partition function via:
\begin{eqnarray}
Z_C(N_B,N_S)&=& 
\int_{0}^{2\pi} \frac{d\varphi_B}{2\pi} e^{-i\varphi_B N_B}
\int_{0}^{2\pi} \frac{d\varphi_S}{2\pi} e^{-i\varphi_S N_S}
Z_{GCR}(\varphi_B,\varphi_S) \, \,.
\label{eq:ZCfourier2d}
\end{eqnarray}

While our data and this formalism allow two conserved charges -- $N_B$ and $N_S$ -- in the present work we will
focus on the $N_B$ dependence only. The study of the role played by strangeness is deferred
to future work.
Hence, we study the baryo-canonical ensemble
\begin{eqnarray}
Z_C(N_B)&=& 
\int_{0}^{2\pi} \frac{d\varphi_B}{2\pi} e^{-i\varphi_B N_B}
Z_{GCR}(\varphi_B,\varphi_S=0)
\label{eq:ZCfourier1d}
\end{eqnarray}
which was also the focus of existing literature.

Notice that we deferred the Fourier transform to after the ensemble average.
Contrary to other works in the literature, we did not introduce 
a canonical determinant. For the simple $\varphi_S=0$ case this is
a complex sequence $\tilde D_N$ for every configuration
\begin{eqnarray}
\tilde D_N(U) = 
\int_0^{2\pi} \frac{d\varphi}{2\pi} e^{-i\varphi N}
D_q(\varphi/3)\,.
\label{eq:candet}
\end{eqnarray}
Since the ensemble average in Eq.~(\ref{eq:ZGCR2})
is linear, it is our freedom, to choose the order: we either 
Fourier transform each determinant function
separately (obtaining $\tilde D_N(U)$) and average these, or
averaging $Z_{GCR}(\varphi)$ and transforming the average once. 
Our choice for the latter is motivated not
merely to save the costs of several million Fourier transforms, but it will
also give us the freedom to introduce the $\alpha$ replica parameter of
Eq.~(\ref{eq:FLTderiv1}), that will play an important role in
section~\ref{sec:grand}.

Here we mention a promising sampling algorithm that is based on 
canonical determinants \cite{Alexandru:2005ix}.
Although the canonical determinants are complex numbers,
$Z_{C}(N)$ is real, as the imaginary parts exactly cancel 
between configuration $U$ and its conjugate $U^\star$.
The real part of $\tilde D_N(U)$ is not positive, though. 
If it was, a straightforward importance sampling could be implemented.
The Kentucky algorithm \cite{Alexandru:2005ix,Li:2010qf}
simulates a canonical ensemble with fixed $N$. It
takes $|\mathrm{Re}\,\tilde D_N(U)|$
instead of $\mathrm{Re}\,\tilde D_N(U)$
as the simulation weight, in a procedure called sign quenching.
In a second step the missing sign is reintroduced by means of reweighting, which is where the sign problem hits.  An analogous procedure was
applied to finite $\mu_B$ in Refs.~\cite{Giordano:2020roi,Borsanyi:2021hbk}.
For staggered fermions the canonical sign quenching has one
advantage over its grand canonical analog:
the staggered rooting is performed on real determinants, which become
complex only after the rooting through the Fourier transform. The Kentucky
algorithm comes with the high costs of computing $\tilde D_N(U)$ at every
update, unlike the approach followed in this work, where we save every 20th
configuration and thus calculate $D_q(\varphi)$ on a much smaller set.

\begin{figure*} 
\centering 
\includegraphics[width=0.32\textwidth]{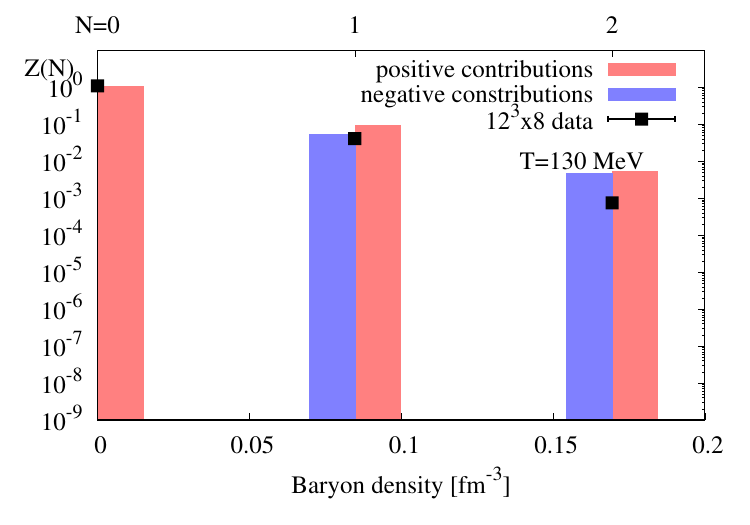}
\includegraphics[width=0.32\textwidth]{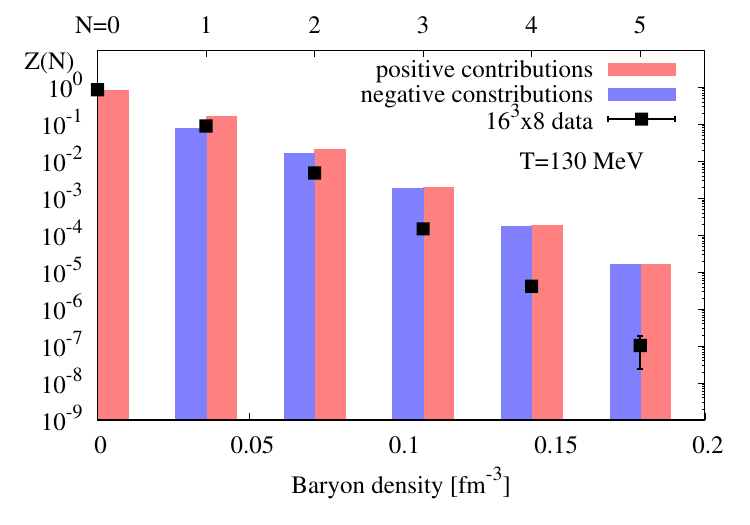}
\includegraphics[width=0.32\textwidth]{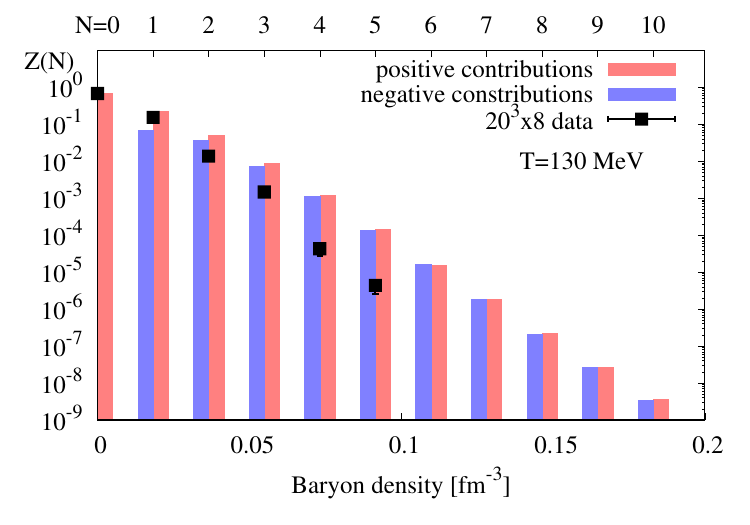}
\caption{\label{fig:signproblem}
We show the canonical partition function as a function of baryon
density for three volumes at a fixed temperature $T=130$~MeV.
The black symbols are the result of the cancellation of positive and
negative contributions, shown as red and blue bars, respectively.
The set of plots illustrates the growing severity of the sign problem
as the density or the volume is increased.
}
\end{figure*}

The sign problem in the canonical ensemble can be illustrated by
displaying the resulting $Z_C(N)$ together with its positive and negative contributions (see Fig.~\ref{fig:signproblem}).
The distance between the top edge of the red bar and the black data point on the
logarithmic scale can be interpreted as the severity of the cancellations
for a given particle number.
Although both $Z(N)$ and the sum of the positive-only contributions
drop near-exponentially with increasing $N$, the rates differ,
resulting in an exponential sign problem.

\subsection{Algorithmic support for the restoration of center symmetry\label{sec:algo:center}} 

Dynamical grand canonical simulations break the $\mathbb{Z}_3$ center symmetry
through the temporal boundary condition, even at $\mu_B=0$. If the
simulation volume is large, this can mean that only a single sector is sampled.
Remarkably, all three center sectors contribute equally to the 
canonical ensemble. The employed lattice algorithm must then ensure
the even sampling of the sectors. This is clearly not automatically
achieved in a grand canonical simulation. For example, with the moderate
volume used here (a $16^3\times8$ lattice), only a small minority of the
configurations belong to either of the two disfavored center sectors.  
See table \ref{tab:stat} for the empirical distribution between center sectors.

\begin{figure*} 
\centering 
\includegraphics[width=0.48\textwidth]{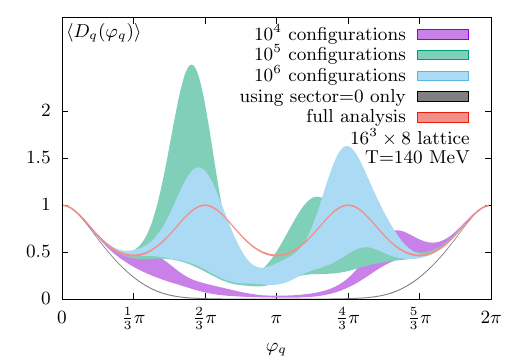}
\includegraphics[width=0.48\textwidth]{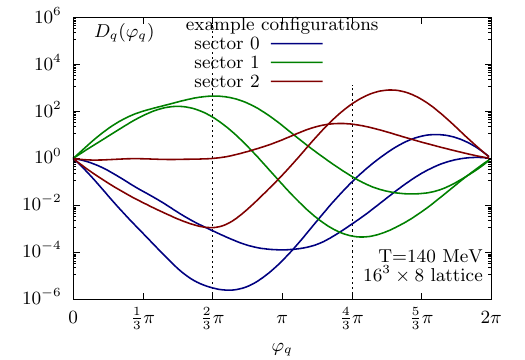}
\caption{\label{fig:center_determinant}
Illustrations of the quark determinant in Eq.~(\ref{eq:Dq}) and our
strategy to compute its ensemble average.
\textit{Left panel:} While the period of $D_q(\varphi_q)$ is $2\pi/3$,
this is very difficult to see even in a high statistics ensemble.
With $10^4$ configurations we get a clearly wrong answer, but even ten
or hundredfold statistics does not bring us much closer to the true
result, either. The solution is the partitioning of the
configuration space into three equivalent center sectors, the zero sector's
result is shown in black ($\langle D_q(\varphi_q)\rangle_{S0}$).
 The red curve is the sum of the black curve
and its two possible shifted forms with an additional normalization
(see Eq.~(\ref{eq:center5})).
\textit{Right panel:}
we show an example configuration for each sector from our pool of configurations.
}
\end{figure*}

The $D_q(\varphi_q)$ of Eq.~(\ref{eq:Dq}) can be computed without
difficulty, and has a period of $2\pi$ by construction.
Upon building an ensemble average 
$\langle D_q(\varphi_q)\rangle=Z_{GCR}(\varphi_q/3)$ has a
period of $2\pi/3$, as required by the center symmetry.
It was already noticed in the earliest canonical work \cite{Hasenfratz:1991ax}
that the naive average $\langle D_q(\varphi_q)\rangle$ contradicts 
this symmetry principle. We illustrate this in Fig.~\ref{fig:center_determinant}.
As Ref.~\cite{Hasenfratz:1991ax} notes, only very good statistics simulations
can recover the correct periodic behavior, shown here in red.
We see in this example that ten-folding the statistics two times in a row
has only made the errors larger, while a precise periodic $\langle D_q(\varphi_q)\rangle$
could not be obtained even in the moderate volume and with the high statistics used in this study. Ref.~\cite{Hasenfratz:1991ax} could identify
the ``outliers'' (shown as sector 1,2 example in the right
panel of Fig.~\ref{fig:center_determinant}).
Ref.~\cite{Hasenfratz:1991ax} also introduces the idea of using 
the geometric mean of rotated versions of the determinant, as a remedy. This, however,
would be equivalent to simulating at an imaginary isospin chemical potential, known to lead to a first order transition
\cite{Yaffe:1982qf,Endrodi:2026tqa}.

In the canonical update algorithms of 
Refs.~\cite{Alexandru:2005ix,Kratochvila:2006jx} an extra sector-changing
update is performed ($U\to U'$), by rotating the temporal link variables
on one fixed time-slices with the center element of the gauge group.
Since the phase on the last time-slice is just how the imaginary-valued chemical is defined, this also means:
\begin{eqnarray}
U \stackrel{e^{-i2\pi/3}}{\longrightarrow} U' \quad: \quad
D_q(U,\varphi_q) \to D_q(U,\varphi+2\pi/3)=D_q(U',\varphi_q)
\end{eqnarray}
In a canonical ensemble $U$ and the center rotated $U'$ and $U''$
configurations have exactly equal weight, given by the arithmetic
(not geometric) mean of rotated and not rotated determinants \cite{Kratochvila:2006jx}.
\begin{eqnarray}
w(U',\varphi_q-2\pi/3)=
w(U,\varphi_q) = 
\frac{1}{3}\left(D_q(\varphi_q)+ D_q(\varphi_q+2\pi/3)+D_q(\varphi_q-2\pi/3
)\right)\,.
\end{eqnarray}

Then, our task is to compute $\langle w(U,\varphi)\rangle$ starting from
a $\mu_B=0$ ensemble.
The key principle for doing so is a partitioning of the configuration
space into three disjoint center sectors. It is then necessary that i) this is well defined for
any given configuration,
ii) a center transformation of an individual configuration 
steps through the sectors, and iii) a center transformation does not change the weight ($w(U,\varphi_q)$) in the canonical ensemble.
A sub-optimal partitioning adhering to these axioms
could be given in terms of the phase of the Polyakov loop,
for the optimal method we ask for the reader's patience.

We will write the ensemble average of an observable $\mathcal{O}$
restricted to one sector, e.g. sector 1 as
\begin{eqnarray}
\left\langle \mathcal{O} \right\rangle_{S1} = 
\frac{
\int \left.DU\right|_{\mathrm{Sector1}} e^{-S_{\rm eff}(U)} 
\mathcal{O}(U)
}{
\int \left.DU\right|_{\mathrm{Sector1}} e^{-S_{\rm eff}(U)} 
}
\label{eq:center1}
\end{eqnarray}
where $S_{\rm eff}$ is the effective action after integrating out 
the quarks at zero chemical potential, where the simulation was run. 
Clearly, the normalization of this path integral differs
from that of the unrestricted integral, because their ratio is not 1/3 due to the $\mu_B=0$ ensemble breaking center symmetry.

Calling the restricted partition functions $Z_{S0}$, $Z_{S1}$ and $Z_{S2}$
for the three sectors, the full expectation value of an arbitrary operator
can be expressed as
\begin{eqnarray}
\langle\mathcal{O}\rangle &=&
\frac{
Z_{S0} \langle \mathcal{O}\rangle_{S0}
+Z_{S1} \langle \mathcal{O}\rangle_{S1}
+Z_{S2} \langle \mathcal{O}\rangle_{S2}
}{Z_{S0} +Z_{S1} +Z_{S2} }
\label{eq:center2}
\end{eqnarray}

Let us now substitute $D_q(\varphi_q)$ for $\mathcal{O}$. The unbroken center symmetry
is manifest in Eq.~(\ref{eq:center2}), and we only need the range $-\pi/3\le \varphi_q \le\pi/3$ for a full
result.
\begin{eqnarray}
\langle D_q(\varphi_q)\rangle &=&
\frac{
 Z_{S0} \langle D_q(\varphi_q)\rangle_{S0}
+Z_{S1} \langle D_q(\varphi_q)\rangle_{S1}
+Z_{S2} \langle D_q(\varphi_q)\rangle_{S2}
}{Z_{S0} +Z_{S1} +Z_{S2} }
\label{eq:center3}
\end{eqnarray}
We exploit that the center transformation ($\varphi_q\to\varphi_q+2\pi/3$)
steps into the next sector, e.g.
\begin{eqnarray}
Z_{S1} \langle D_q(\varphi_q)\rangle_{S1} &=& 
\int \left.DU\right|_{\mathrm{Sector1}} e^{-S_{\rm eff}(U)} 
D_q(\varphi_q)\nonumber\\
&=&\int \left.DU\right|_{\mathrm{Sector0}} e^{-S_{\rm eff}(U)} 
D_q(\varphi_q+2\pi/3)
=
Z_{S0} \langle D_q(\varphi_q+2\pi/3)\rangle_{S0} 
\label{eq:center4}
\end{eqnarray}
\begin{eqnarray}
\langle D_q(\varphi_q)\rangle &=&
\frac{
 Z_{S0} \langle D_q(\varphi_q)\rangle_{S0}
+Z_{S0} \langle D_q(\varphi_q+2\pi/3)\rangle_{S0}
+Z_{S0} \langle D_q(\varphi_q-2\pi/3)\rangle_{S0}
}{Z_{S0} +Z_{S1} +Z_{S2} }
\label{eq:center5}
\end{eqnarray}
Using Eq.~(\ref{eq:center4}) also to relate $Z_{S0}$, $Z_{S1}$ and $Z_{S2}$
as 
\begin{eqnarray}
\frac{Z_{S1,S2}}{Z_{S0}} &=& \langle D_q(\pm2\pi/3)\rangle_{S0}
\label{eq:center6}
\end{eqnarray}
we finally write the desired expectation value solely employing sector 0 configurations:
\begin{eqnarray}
\langle D_q(\varphi_q)\rangle &=&
\frac{
  \langle D_q(\varphi_q)\rangle_{S0}
+ \langle D_q(\varphi_q+2\pi/3)\rangle_{S0}
+ \langle D_q(\varphi_q-2\pi/3)\rangle_{S0}
}{
  \langle D_q(0)\rangle_{S0}
+ \langle D_q(+2\pi/3)\rangle_{S0}
+ \langle D_q(-2\pi/3)\rangle_{S0}
} \, \,.
\label{eq:center5}
\end{eqnarray}
For an optimal partitioning of the configuration space into center
sectors we should have a second look at Fig.~\ref{fig:center_determinant}.
What makes the naive averaging of $D_q(\varphi_q)$ difficult between
$\pi/3$ and $5\pi/3$ is the dynamical range of the determinants.
The six random examples in Fig.~\ref{fig:center_determinant}
spread across nine orders of magnitude, which would be even more in a larger volume.
The true expectation value of $\langle D_q(2\pi/3)\rangle =1$ is
realized as an average including some extremely rare ($\mathcal{O}(10^{-3})$)
configurations in sectors 1 and 2 with a contribution of
($\mathcal{O}(10^{+3})$).

Every configuration can be sampled by either of the three ensembles,
defined at $\mu_{q}/T = i \phi_q \in \{0, i 2\pi/3, -i2\pi/3\}$, though
with different probability. It is the determinant ratio $D_q(U,\varphi_q)$
itself that tells this relative probability: a configuration $U$ will
be generated in an ensemble at $\mu_q/T=i 2\pi/3$ with a factor
$D_q(U,2\pi/3)/D_q(U,0)$ larger probability in comparison to the ensemble
at $\mu_q=0$. Our partitioning is based on the chemical potential
(one of $\{0, i 2\pi/3, -i2\pi/3\}$) where a given configuration 
has the highest probability to appear. In other words, whichever of the
three positive (rooted) determinants is the largest,
\begin{eqnarray}
D_q(U,0),\quad D_q(U,2\pi/3),\quad D_q(U,-2\pi/3)
\end{eqnarray}
specifies the sector to be S0, S1 or S2, respectively. This criterion trivially
fulfills the axioms required above and is optimal in the sense that 
it keeps only those configurations from the $\mu_B=0$ ensemble, that
would not have been better sampled by other ensembles, such as $\mu_q=\pm i2\pi/3$. With this selection rule, Eq.~(\ref{eq:center5}) gives a very precise
result. We show this in red in Fig.~\ref{fig:center_determinant}.
The percentage of the configurations that contribute to the final
average, and the discarded ones in sector 1 and 2 are listed
in Table~\ref{tab:stat}.



\section{Thermodynamics at finite density\label{sec:thermo}}

The observable that we directly computed on the lattice in the previous
section is $Z_{GCR}(T,V,\varphi)$, which gives the canonical partition function $Z_C(T,V,N)$ through a Fourier transform in 
$\varphi$ as written in Eq.~(\ref{eq:ZCfourier}).
The limiting factor of how large an $N$ we can extract with this Fourier transform is the
statistical error. 

\begin{figure}
\centering
\includegraphics[width=0.48\textwidth]{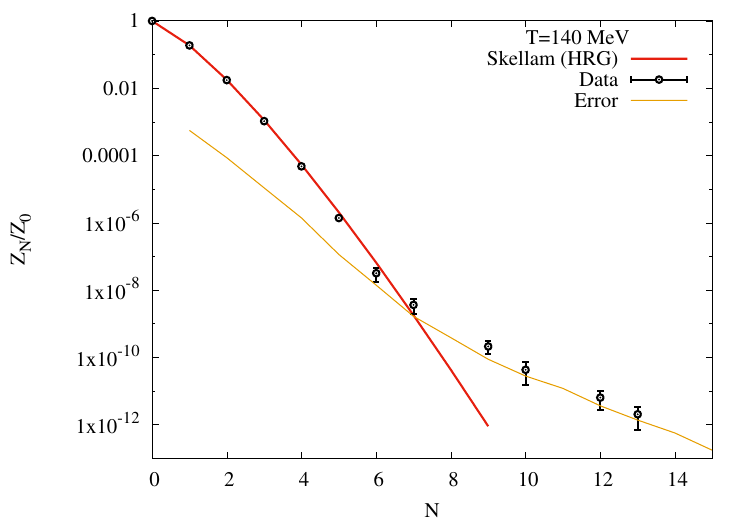}
\caption{\label{fig:ZB140}
The canonical partition sum $Z_C$ for the example of $T=140$~MeV.
For the lowest particle numbers we find good agreement with
the Skellam distribution. The level of the statistical error
is shown as an orange line. From $N=6$ on the statistical error
is 100\%.  The magnitude of $Z_N$ then simply follows the error
as a function of $N$.
}
\end{figure}

An example of the resulting $Z_C(T,V,N)$ is shown in Fig.~\ref{fig:ZB140} at
$T=140$~MeV, which is the temperature with the highest statistics (see
Table~\ref{tab:stat}). Sitting below the pseudo-critical temperature, we expect
$Z_C(N)$ to follow the Skellam distribution, which we show as a red curve
in Fig.~\ref{fig:ZB140}. The parameter of the 
Skellam distribution (see $\hat p_B$ in Eq.~(\ref{eq:cosh}))
is the leading term in the fugacity expansion that we extracted from the grand canonical data, not by fitting
the points in Fig.~\ref{fig:ZB140}.

The agreement with data is good, though the statistical error (orange line)
decreases with $N$ more slowly than the signal, eventually resulting in 100\% errors for $N\ge6$
in this case. The apparent slope of noise-dominated data reflects the error's
scaling with $N$, and the algorithm does not protect us from
unphysical, negative values ($N=8$). 
In our analysis we will discard the marginally significant data points
based on the error on $Z_N$. 

\begin{figure*}
\centering
\includegraphics[width=0.48\textwidth]{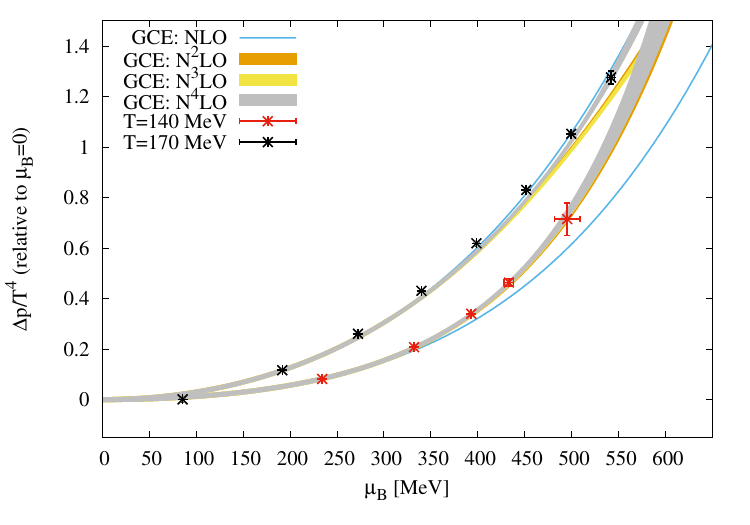}
\includegraphics[width=0.48\textwidth]{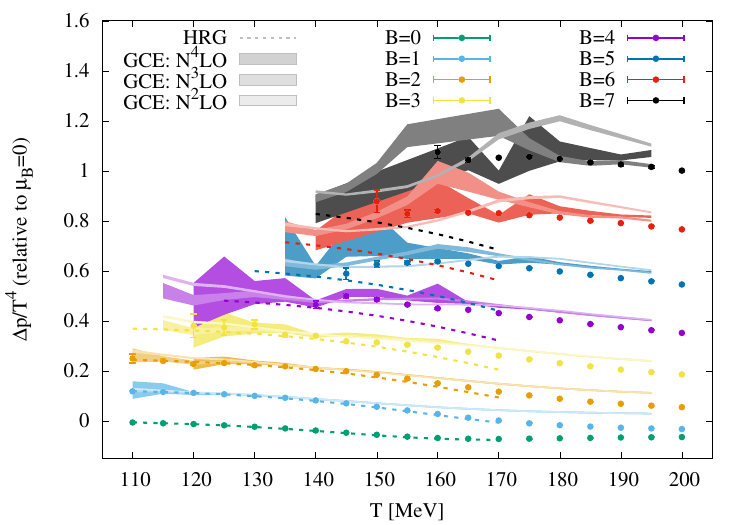}
\caption{\label{fig:canondp}
Left: The relative pressure at two temperatures in the canonical ensemble.
We show the Taylor expanded
grand canonical results (based on the same ensemble) as color
bands.
Right: The relative pressure as a function of the temperature 
at fixed baryon number. We also show the grand canonical extrapolation as
bands and the prediction of the HRG model as dotted lines 
}
\end{figure*}

From $Z_C(T,V,N)$ the computation of $\mu_B$ is straightforward,
using the backward discrete derivative of Eq.~(\ref{eq:mub}), while the canonical
pressure requires a separate Fourier transformation (see
Eq.~(\ref{eq:dpcanon})). The relative pressure that can be readily
extracted from our data employs a hybrid definition: it subtracts
the grand canonical pressure at $\mu_B=0$ from
the canonical pressure with particle number $N$, with this difference possibly
negative, especially for $N=0$.
We show the relative pressure according to this definition in Fig.~\ref{fig:canondp}. In the left panel we plot the result of 
Eq.~(\ref{eq:dpcanon}) against Eq.~(\ref{eq:mub}) together with the Taylor extrapolation to four consecutive orders in $\mu_B^2$.

Throughout this work we will use the 'GCE' label to refer to the
Taylor extrapolation using the very same lattice ensembles. Leading order 
means that the expansion is truncated at the order of $\chi^B_2(T)$,
and 
the N$^4$LO order includes up to $\chi^B_{10}(T)$. These coefficients
have already been published in Ref.~\cite{Adam:2025hpb}.

In order to determine the chemical potential, the Taylor expanded result for $n_B(\mu_B)$ was computed to the chosen
order, then inverted to translate the density $N/V$ to 
$\mu_B$.
This was not always successful, since high order fluctuations are
notoriously noisy and the error on Taylor expanded results is often
higher than on canonical results, even though both approaches use the
very same set of configurations. One possible reason is a potential cancellation between the noise in higher orders of the Taylor expansion in a resummed setting, like the canonical ensemble. Further, the algorithmic improvement of section \ref{sec:algo:center} has not yet been applied to the Taylor coefficients.
How many orders in the Taylor expansion
are needed for a given accuracy may only be known empirically, and even that
is subject to statistical error. Adding one further order might perhaps only increase the error, 
but it can just as well shift the value enough to be incompatible with the previous order.

In the right panel of Fig.~\ref{fig:canondp} we show a comparison of our canonical results 
with the three highest orders in the GCE. Especially for higher densities, subsequent orders are not in agreement. Our canonical data, which are compatible
with the highest order except at high temperature, have consistently smaller errors.
We note that, especially at high $T$, the volume is small and
the canonical and grand canonical observables are expected to agree in the
thermodynamical limit only. We also show a canonical prediction from the hadron resonance gas (HRG) model
in the same figure, computed by Fourier transforming
the exponentiated grand canonical potential within HRG using the simulation
volume $8/T^3$ employed in this work. We could crosscheck the canonical HRG thanks to the Thermal-FIST package \cite{Vovchenko:2019pjl}. HRG is expected to describe
data in the hadronic phase. As the density is increased, the range of agreement
shifts to smaller temperatures, consistent with a density-dependent transition
temperature. We find that at fixed density HRG underestimates the pressure difference,
which is expected for a repulsive two-baryon interaction. Note that the same
interactions in turn cause HRG to overestimate $\Delta p/T^4$ if $\mu_B$ is kept fixed,
instead.
The negative relative pressure at $N=0$ is also apparent in the right
panel of Fig.~\ref{fig:canondp}. Here, the fully canonical relative pressure
would be the difference between the $N>0$ and the $N=0$ results.

\begin{figure}
\centering
\includegraphics[width=0.48\textwidth]{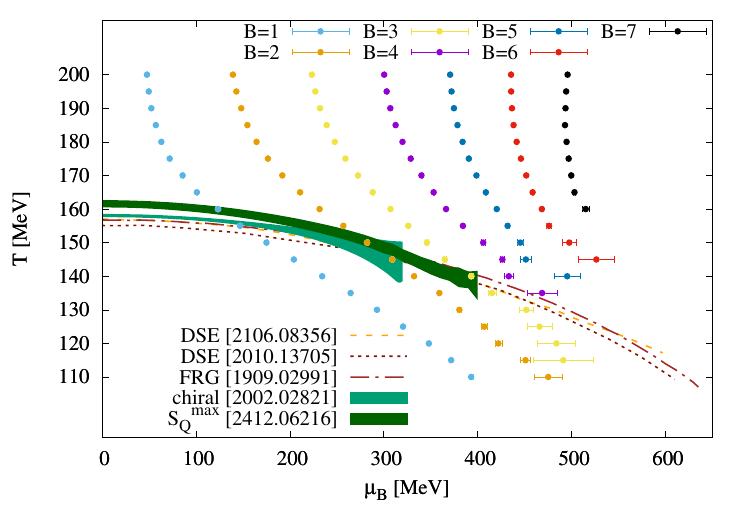}
\caption{\label{fig:can_phase_diag}
Contours of constant baryon number in the simulation volume $V=8/T^{3}$
on the QCD phase diagram.
}
\end{figure}

In Fig.~\ref{fig:can_phase_diag} we plot the chemical potential
determined as the discrete (backward) derivative of the Helmholtz free
energy for a set of net baryon numbers. We use the standard axes
as in the QCD phase diagram, and show the QCD crossover lines from Refs.~\cite{Gunkel:2021oya,Borsanyi:2020fev,Borsanyi:2024xrx}.  We note that the chiral crossover
line of \cite{Borsanyi:2020fev} is continuum extrapolated and belongs to the strangeness
neutral case, whereas our canonical data come from a finite lattice spacing
and a small volume $(16^3\times8)$, and the strangeness chemical potential
is set to zero for simplicity $\mu_S=0$. In Ref.~\cite{Borsanyi:2024xrx} 
we estimated the transition line from the static quark entropy (related
to the Polyakov loop) in the $\mu_S=0$ scheme \cite{Borsanyi:2024xrx}, employing the very same ensembles as we used here with updated statistics.

\section{Generalized canonical results and the grand canonical limit\label{sec:grand}}

The goal of this section is to compute predictions at real chemical potentials
in the grand canonical context, starting from our canonical setup. In fact, Eq.~(\ref{eq:ZGCrel}) shows this connection, and the same equation has already been used in
this context~\cite{Danzer:2012vw}. Unfortunately, we do not know $Z_C(T,V,N)$
for high $N$, hence the use of Eq.~(\ref{eq:ZGCrel}) is not practical if we are
interested in a truncation-free result. Instead, we don't start from $Z_C(T,V,N)$, but from one step before: the  ratio $Z_{GCR}(T,V,\varphi)$, which we can access directly on the lattice.

In section \ref{sec:thermo} we defined all observables starting from the free energy
\begin{eqnarray}
F(T,V,N) &=& -T \log\int_0^{2\pi} \frac{d\varphi}{2\pi} e^{-i\varphi N + \log Z_{GCR}(T,V,\varphi)}\,.
\end{eqnarray}

We now generalize this relation with the replica parameter ($\alpha$) of 
Eq.~(\ref{eq:FLTderiv1}) as 
\begin{eqnarray}
F_\alpha(T,\alpha V,N) &=& -\frac{T}{\alpha} \log\int_0^{2\pi} \frac{d\varphi}{2\pi}
e^{-i\varphi N + \alpha\log Z_{GCR}(T,V,\varphi)}
\label{eq:Falpha}\\
N&=& \alpha n V\,,
\label{eq:Nalpha}
\end{eqnarray}
where $n$ is the net baryon density. The parameter $N$ must be an integer,
because of the manifest periodicity of the integrand. On the other hand, the density
$n$ is not limited to integer multiples of $1/V$, and any $n$ can be realized
by fixing $N$ and $\alpha$ according to Eq.~(\ref{eq:Nalpha}).
The definition of Eq.~\eqref{eq:Falpha} can be easily related to the
asymptotic formula in Eq.~(\ref{eq:FLTasymp}) by
\begin{eqnarray}
F_\alpha(T,\alpha V,N) &=& F_{LT}(T,V,n V)
-\frac{T}{2}\alpha^{-1}\log{\alpha^{-1}}
+\mathcal{O}(\alpha^{-1})
\label{eq:Falphaasymp}
\end{eqnarray}
where the asymptotic form is valid for fixed density $n$.
We can conveniently use our direct lattice result on $Z_{GCR}(T,V,\varphi)$ to evaluate Eq.~(\ref{eq:Falpha}) for an arbitrary positive 
$\alpha$ parameter.

The physical meaning of e.g $\alpha=2$ is that we have two finite volume
systems, which combine to having $N$ (net) baryons. Although
non-integer values of $\alpha$ do not share this intuitive interpretation, Eq.~(\ref{eq:Falpha}) admits any $\alpha>0$.
In another analogy, $\alpha$ rescales the volume of the 
grand canonical ensemble, before it is converted to canonical. 
However, this picture is valid only for asymptotically large volumes
(e.g. $L\gg m_{\pi}^{-1}$),
where the grand potential $\Omega$ is truly a homogeneous function of $V$.
Of course, $\alpha$ does not change the real system size, or how many
inverse pion masses fit in the simulated box.
Independently of the physical interpretation of the $\alpha$ parameter,
we will treat it as a formal tool to connect the lattice data
at imaginary $\mu_B$ to the grand canonical ensemble at real $\mu_B$.

We propose to use Eq.~(\ref{eq:Falphaasymp}) to take the 
$\alpha^{-1}\to0$ limit numerically on the simulation data with
fixed density and obtain
\begin{eqnarray}
\Omega(T,V,\mu_B)+nV\mu_B &=& 
F_{LT}(T,V,nV) = \lim_{\alpha\to\infty} F_\alpha(T,\alpha V,nV\alpha) \,\,,
\end{eqnarray}
as a result of the grand canonical limit. This translates on
the level of observables to
\begin{eqnarray}
\mu_B(T,V,n) &=&  F_\alpha(T,\alpha V,nV\alpha) - F_\alpha(T,\alpha V,nV\alpha-1)
+ \mathcal{O}(\alpha^{-1})\,,
\label{eq:mualpha}\\
\Delta p_C(T,V,n) &=&
\frac{T}{ V}
\frac{
\int_0^{2\pi} \frac{d\varphi}{2\pi}
\left[Z_{GCR}(\varphi)\right]^\alpha
 \log Z_{GCR}(\varphi) e^{-i\varphi N}
}{
\int_0^{2\pi} \frac{d\varphi}{2\pi}
\left[Z_{GCR}(\varphi)\right]^\alpha
 e^{-i\varphi N }
}
+ \mathcal{O}(\alpha^{-1})\,.
\label{eq:dpalpha}\\
\frac{1}{\chi_2(T,V,n)} &=&VT^2\left(
F_\alpha(T,\alpha V,nV\alpha+1) - 2F_\alpha(T,\alpha V,nV\alpha-1)
+ F_\alpha(T,\alpha V,nV\alpha-1)\right)
+ \mathcal{O}(\alpha^{-1})\,.
\label{eq:chi2alpha}
\end{eqnarray}

\begin{figure*}
\centering
\includegraphics[width=0.48\textwidth]{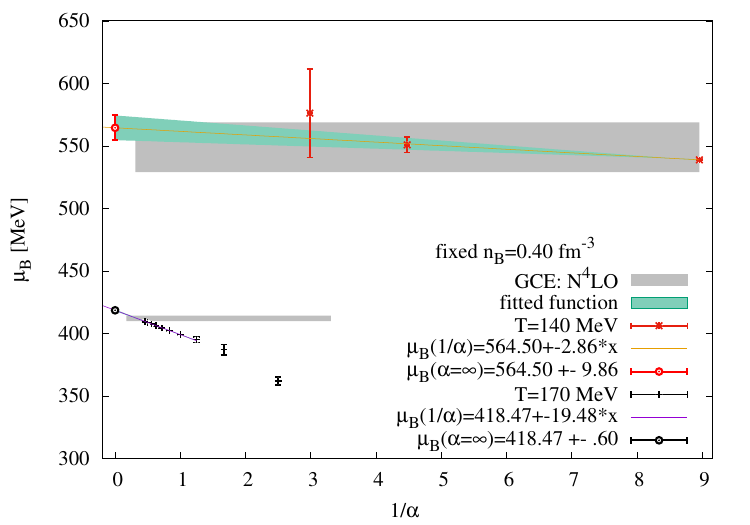}
\includegraphics[width=0.48\textwidth]{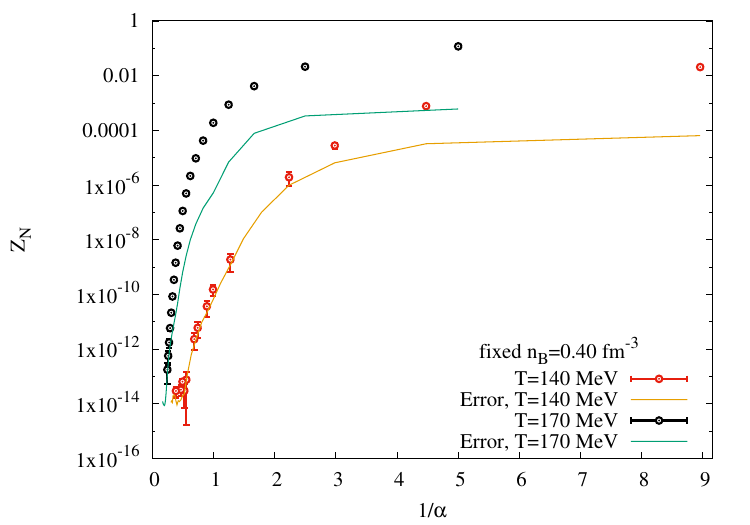}
\caption{\label{fig:alphalimit}
Left: The baryo-chemical potential as a function of $1/\alpha$ and the
$\alpha^{-1}\to0$ limit based on simulation data. We show two temperatures
at fixed baryon density. The grey band shows the Taylor-extrapolated grand
canonical result using coefficients up to $\chi^B_{10}$ \cite{Adam:2025hpb}.
Right: The underlying $Z_N$ is shown on a logarithmic scale at fixed density.
The solid lines indicate the $1\sigma$ error level. For large $N$ (small
$1/\alpha$) the signal is lost, limiting the range of $1/\alpha$ usable for the extrapolation.
}
\end{figure*}

Since these quantities can be computed at fixed density with
various integer $N$ (by selecting $\alpha=N/(nV)$), a limit
procedure in $1/\alpha \sim 1/N$ can be defined. The range in $N$
is limited by the sign problem (especially at low temperature) and
numerical precision. Still, at small $N$ higher order corrections
(e.g. $\sim 1/N^2$) can be relevant, for example at higher temperatures
where the higher precision makes them appreciable.
We illustrate the limit procedure of Eqs.~(\ref{eq:mualpha})-(\ref{eq:dpalpha})
on our simulation data at two temperatures in Fig.~\ref{fig:alphalimit}.

\begin{figure*}
\centering
\includegraphics[width=0.48\textwidth]{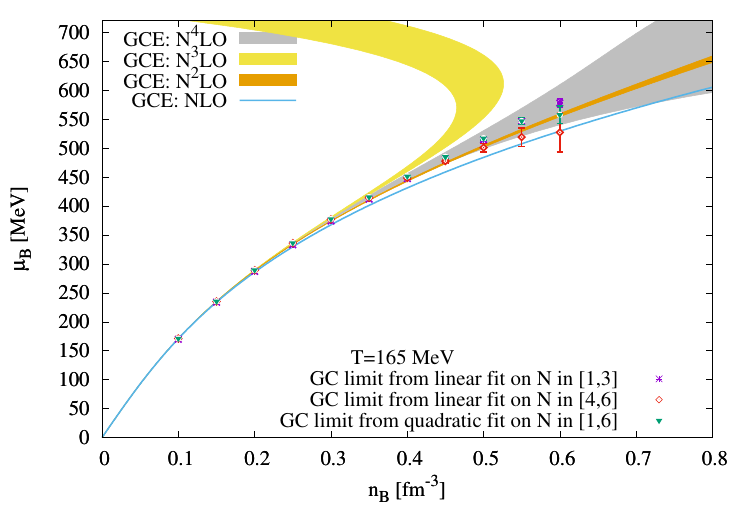}
\includegraphics[width=0.48\textwidth]{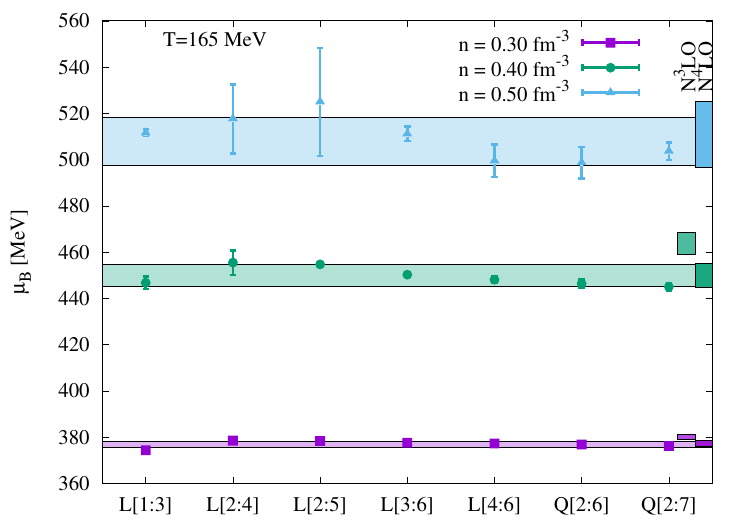}
\caption{\label{fig:muB_fit_syst}
We illustrate the systematic effects of the extrapolation in $1/\alpha$.
In this example ($T=165$~MeV) we show the baryo-chemical potential as a function
of the net-baryon density and compare the results of the $1/\alpha$ extrapolation to various orders (error bars) of the Taylor expansion. In the left panel we show this as a function of temperature. The right panel shows the results of
various linear (L) or quadratic (Q) fits in $1/N$ in the given ranges. The horizontal band shows the combined error
from this procedure, while the bars in the right edge indicate two consecutive orders of the Taylor expansion.
}
\end{figure*}

The range of $1/\alpha$ (or $1/N$) that can be used in 
Eqs.~(\ref{eq:mualpha}-\ref{eq:dpalpha}) to perform the grand canonical
limit ($\alpha^{-1}\to0$) depends on data quality and the size
of higher order terms. This is very similar to a continuum limit $a^2\to0$, which can be reached only in the form of an extrapolation.
Fig.~\ref{fig:muB_fit_syst} presents various versions of this grand canonical
limit, depending on the fitting range and the order in $1/N$, which mostly agree within $1\sigma$. Below $T<165$~MeV we do not have the luxury
to vary the fitting range, e.g. at the lower temperature in Fig.~\ref{fig:alphalimit} only three points are available for this extrapolation.

\begin{figure}
    \includegraphics[width=0.48\textwidth]{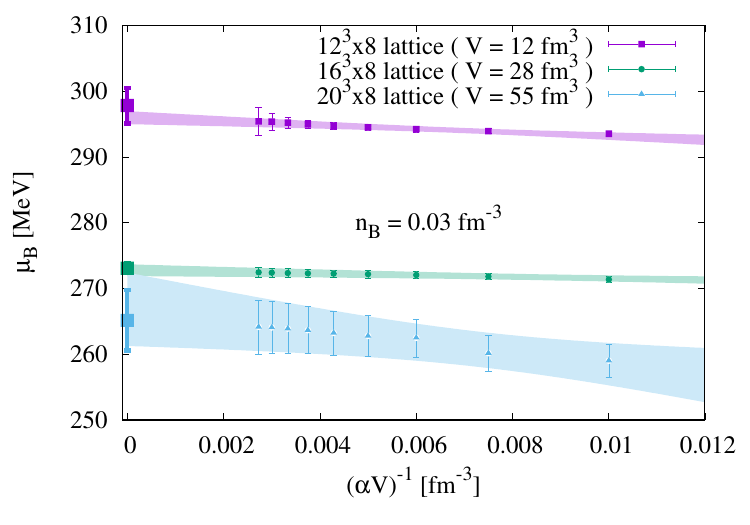}
    \caption{
    \label{fig:volume}
    The lattice data and the corresponding $1/\alpha\to0$ limit at fixed density for three different simulation volumes. The different limits reveal the residual volume dependence of the original grand canonical setup.
    The error bars on the $\mu_B$ axis show the N$^4$LO Taylor expanded grand canonical result ($T=130$~MeV).
    }
\end{figure}

On the other hand, we are able to compare the results of this procedure on three different simulation volumes, namely
$\lat{12}{8}$, $\lat{16}{8}$ and $\lat{20}{8}$, as we show in Fig.~\ref{fig:volume}. As expected, we observe the
same linear approach to the GCE, and also see that the two larger volumes agree within error.

\begin{figure*}
\centering
\includegraphics[width=0.45\textwidth]{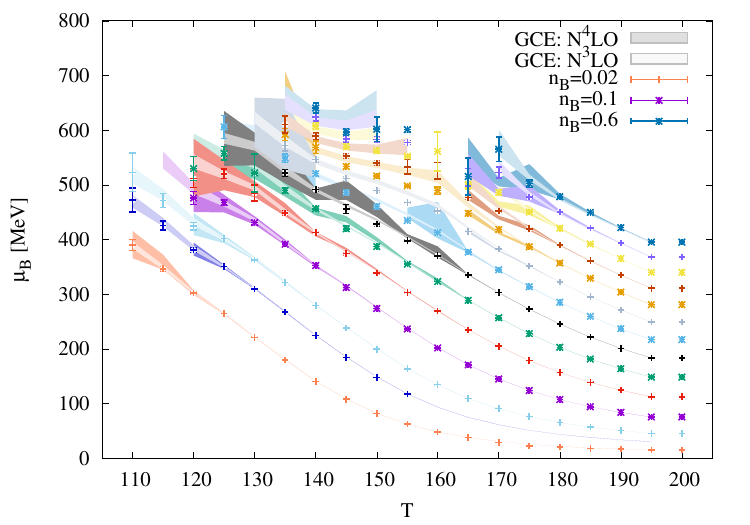}
\includegraphics[width=0.45\textwidth]{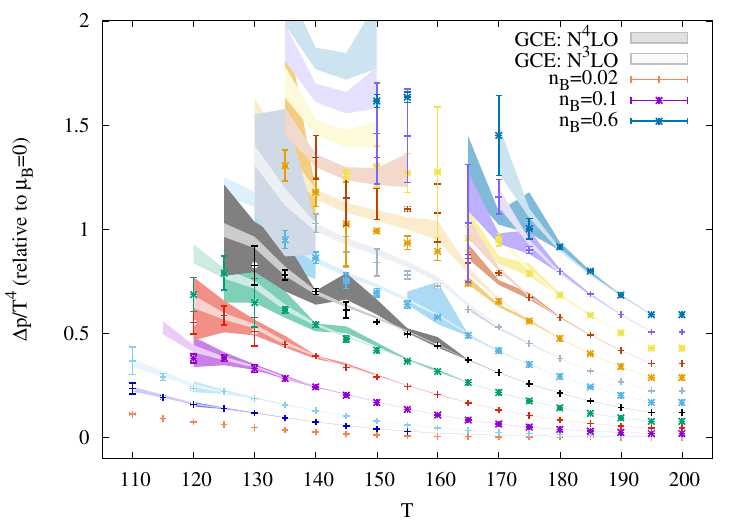}
\caption{\label{fig:obs_inf_alpha}
Infinite $\alpha$ limit of $\mu_B(T,V,n)$ and  $\Delta p_C(T,V,n)$
(using Eqs.~(\ref{eq:mualpha}) and  (\ref{eq:dpalpha}), respectively).
For $\mu_B$ and $\Delta p_C$ each color represents a density $n_B$, with the same coloring scheme as in Fig.~\ref{fig:can_phase_diag_inf_alpha}.
}
\end{figure*}

For our final results in Fig.~\ref{fig:can_phase_diag_inf_alpha} 
and Fig.~\ref{fig:obs_inf_alpha}, we use this range $N=[1,3]$ below $T<165$~MeV,
and the range $N=[4,6]$ for $T\ge165$~MeV. 
Since this result is not continuum extrapolated,
we do not elaborate on the precise magnitude of the systematic
errors.

In Fig.~\ref{fig:obs_inf_alpha} we show the baryo-chemical potential (left) and the relative pressure (right) at a set of fixed
baryon densities ($n_B=$0.02, 0.04, 0.06, then 0.10$-0.60~\mathrm{fm}^{-3}$ in steps of $0.05~\mathrm{fm}^{-3}$), and compare our results
with the Taylor expansion. Note that here also the data points represent
grand canonical results, since the $\alpha\to\infty$ limit has already
been taken. It is natural that, for large densities, the data points are less controlled due to the combined effect of the sign and 
overlap problems. The Taylor expansion seems to have a more severe problem near 
$T=150~$MeV, although both the Taylor expanded grand canonical and the
$\alpha$-extrapolated canonical data are based on the same set of
configurations.

Let us finally turn to the baryon susceptibility $\chi_2(T)$. Eq.~\eqref{eq:chi2alpha} can, in principle, be used to perform an extrapolation, similar
to those for $\mu_B$ and $\Delta p_C$. However, the range of $N$ for a given density
where a linear fit in $1/N$ would be practical is often fewer than three data points.
A much simpler strategy can be used, instead: the function $\mu_B(n_B)$
of Eq.~(\ref{eq:mualpha}) can be derived to yield $\chi_2^{-1}=T^2 d\mu_B/dn_B$.
For this, we fix the range in $N$ and compute $\mu_B(n_B)$ up to the point where $\mu_B$
reaches the maximal value of interest (here 500~MeV). This function can then be derived
with respect to the density ($n_B$) numerically without any problem. 

\begin{figure*}
\centering
\includegraphics[width=0.45\textwidth]{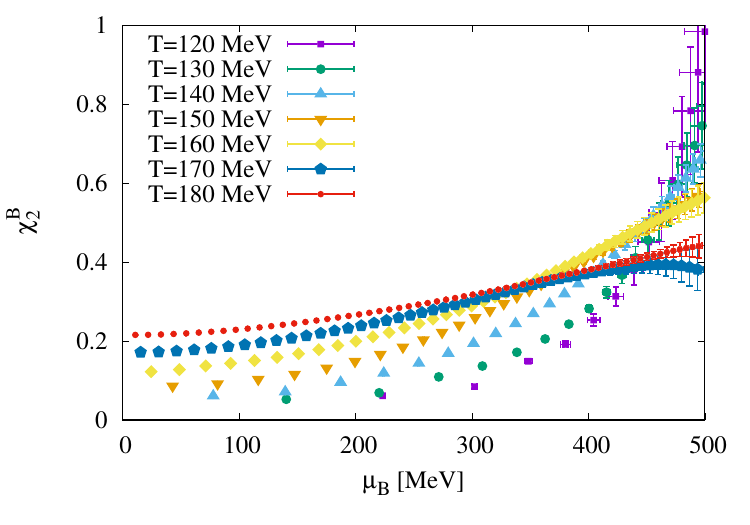}
\includegraphics[width=0.45\textwidth]{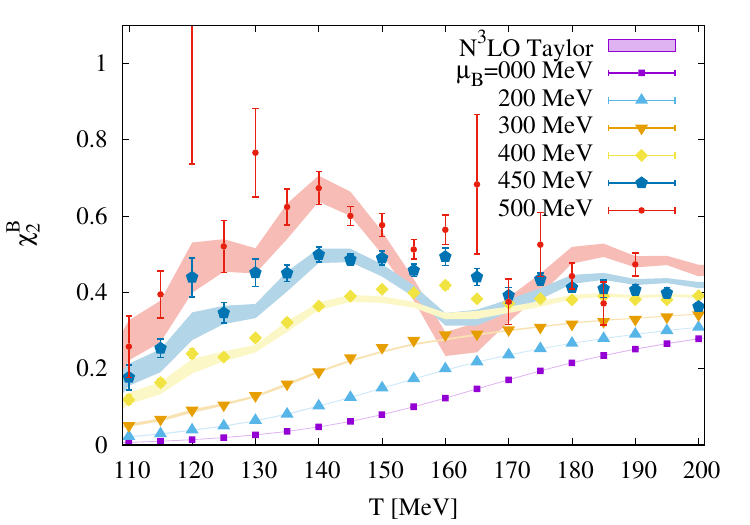}
\caption{\label{fig:chi2}
The net-baryon susceptibility as a function of chemical potential (left), or temperature (right). $\chi_2$ was obtained as
the inverse derivative of the $\mu_B(n_B)$ function. The Taylor extrapolation in the right panel uses $\chi^B_8$ as its highest order, while using $\chi^B_{10}$ would result in much higher errors.
Notice, that $\chi^B_2(T)$ starts out as a sigmoid for small $\mu_B$ but develops as $\mu_B$ is increased.
}
\end{figure*}

In Fig.~\ref{fig:chi2} we show the resulting baryon susceptibility. The left
panel shows the $\mu_B$ dependence of this inverse derivative for a selection of
temperatures. In the right panel we interpolate every jackknife sample to a given
fixed $\mu_B$ so that the error bars are on $n_B$ or $\chi_2$, not on $\mu_B$. Through
this interpolation we finally return to a completely grand canonical picture.
In the right panel we also show the Taylor extrapolated prediction, using the
coefficients at $\mu_B=0$ up to $\chi^B_8$.
Both the Taylor extrapolation and the data points, 
i.e. the grand canonical result we obtained though the canonical detour, are from the same
set of input configurations, and different temperatures are statistically independent.
In the figure we observe that a) the Taylor extrapolation introduces a non-physical
non-monotonic temperature dependence, which has already been discussed in the
context of the $T'$ expansion \cite{Borsanyi:2021sxv}; b) the data points
reveal that for small $\mu_B$ one recovers the known sigmoid form of $\chi_2(T)$. 
As $\mu_B$ is increased, the inflection point moves to smaller temperatures,
as expected. For even larger chemical potential a new maximum develops.

The signal quickly deteriorates as the chemical potential is increased. Moreover,
some data points (e.g. those at $T=120$ and 130~MeV) give higher fluctuations than the
neighboring temperatures. Such outliers can be hints for the presence of an overlap
problem. Indeed, in this work we have not addressed the overlap nor the
sign problem of reweighting. What we did is to use the canonical formalism
and derive a new method to extract finite density physics from $\mu_B=0$ ensembles,
that eliminates some of the shortcomings of reweighting or the Taylor expansion.

\section{Conclusions and outlook\label{sec:prospects}}

In this work, we used lattice QCD simulations at $\mu_B=0$ to extract physical observables
(baryon density, pressure, fluctuations) at real $\mu_B>0$.
Similar results have already been published  with various extrapolation
schemes for more than a decade.
Specifically, the Taylor expansion has been
used extensively to compute the very same observables we presented here. 
The computation of the Taylor coefficients is not hindered by the complex action
problem, but it still suffers from a remnant sign problem \cite{Adam:2025hpb},
making the simulation of higher order coefficients extremely costly, unless
the volume is small. Even in a small volume, it is difficult to know
how well a fixed order approximates the true result. Adding one more order can
be impossible due to its cost, or because simulations deliver results for the highest order with insufficient precision due to the limited statistics.
It is also known that the use of imaginary-valued $\mu_B$ can boost the range
of such extrapolations. 
In general, the numerical cost of Taylor coefficients
is driven by the severity of the cancellation problem, which in turn is proportional to a power of the volume (the exponent depends on the order). Thus, it is generally more cost effective to model the
imaginary-$\mu_B$ dependence of lower-order susceptibilities than to address directly high order fluctuation observables~\cite{DElia:2016jqh,Borsanyi:2018grb}.
Yet, the extrapolation itself is ill-posed, and is subject to
hard-to-control systematic errors. The $T'$ expansion \cite{Borsanyi:2021sxv}
introduced a more physical ground for the $\mu_B$ extrapolation where the lowest
orders dominate at moderate $\mu_B$. 

On the other hand, this work \textit{extracts}, instead of extrapolating, finite density information.
In this sense our method is  similar to reweighting with one important technical difference: the simple reweighting
cannot be applied to rooted actions, like staggered lattice QCD. While other formulations,
e.g. Wilson, twisted mass or minimally doubled fermions might in principle be reweighted
from zero to real $\mu_B$, with these
actions one struggles to simulate with physical quark masses even at $\mu_B=0$.
As of today, the only lattice formulation where the extreme statistics needed to fight
off the sign problem is not prohibitively expensive, is the staggered quark.

Our method can be summarized with three steps:
1) We first reweighted the $\mu_B=0$ ensembles to imaginary chemical potential, then 2) moved to the canonical ensemble by means of a Fourier transform, and finally 3) performed
a limit procedure to return to the grand canonical ensemble at finite $\mu_B$.

Already step 1) required a new algorithmic ingredient, in order to ensure a proper sampling
of the center sectors in the center-breaking setup at $\mu_B=0$. This improvement
might also turn out to be beneficial for the computation of the Taylor coefficients, 
as it effectively eliminates ill-sampled parts of the configuration space. These eliminated parts are not left out of the path integral, but computed precisely
through an exact transformation of a well-sampled sector. In this work we did
not use simulations at imaginary valued $\mu_B$, but an extension of the method
can be easily derived. 
Quite importantly, not the used range of imaginary chemical
potentials should be assigned to various imaginary-$\mu_B$ ensembles, but the
configuration space has to be partitioned. We leave this extension to future
work.

Step 2) has been preformed by several authors before us, although never with physical quark masses. We did not address stability and precision, because in most cases the sign
problem, stronger in our system than in other literature, was more significant.
In the volume of our choice ($LT=2$) the sign problem is already out of control in the presence of a handful baryons, corresponding to a large density. We demonstrated the magnitude of the sign problem in three simulation volumes.

Leaving the result in the canonical formulation would severely limit its usability.
We computed the pressure and the chemical potential for just a handful of densities
where $N$ is an integer. The volume matters, too, since canonical effects are very
different in our simulation volume and in the entire volume of the quark gluon plasma
in a heavy ion collision event.

Thus, we introduced step 3) to recover the grand canonical description. Instead of naively
adding up the canonical partition function with weights $e^{N\mu/T}$ we do employed a limit procedure with a known asymptotic form. We determined the leading term of
this asymptotic form for a non-interacting hadron gas (HRG in the Boltzmann
approximation). This limit procedure introduces systematic errors that we studied
on a few examples, but did not further elaborate. Moreover, we worked at finite lattice
spacing and, most importantly, in a volume where the sign problem can be kept
under control up to $\mu_B\le 500$~MeV.

We focused on bulk QCD thermodynamics, charting the QCD phase diagram 
and computing the pressure and its $\mu_B$ derivatives, from which the 
full equation of state can be determined.
The results presented here can be extended to account for more than one
conserved charge without technical difficulties. The results we presented were defined at $\mu_S=0$, but an extension to strangeness neutrality can be a natural
continuation of this work. It should also be investigated whether the canonical
method and reweighting agree when a real isospin chemical potential is introduced. In the latter case it is known that
the Taylor extrapolation is limited by the finite radius of convergence of $m_\pi$ due to the transition to a pion-condensed phase~\cite{Son:2000xc}.

In this work we employed a finite lattice spacing ($N_t=8$). Fortunately, the severity
of the sign problem does not diverge towards the continuum limit. However, a technical
difficulty is the extraction of the eigenvalues from the simulated configurations.
This step scales as $\sim N_t^9$, and its numerical stability is
uncertain at $N_t=12$. We suspect that in such a case a stochastic algorithm would probably be better suited.

The biggest challenge is to increase the simulation volume, which would be a welcome extension. This is mainly because a larger volume would give us access to chiral observables, which are distorted in the volume we employ here.
For example, at a chemical potential $\mu_B\approx 500~\textrm{MeV}$
we have ca. 5 net-baryons in the $16^3\times 8$ lattice volume. 
A lattice size of $20^3\times8$ corresponds to doubling the physical volume and having
10 net-baryons.
The corresponding partition function (the signal)
is then $Z_{16^3\times8}(N=5)\sim 10^{-7}$ and
$Z_{20^3\times8}(N=10)\sim 10^{-14}$, respectively. The signal is,
thus, suppressed by a factor of 160 on the smaller lattice, and by a factor
of $3.6\cdot10^5$ on the larger one. If one does not increase the statistics while
doubling the volume, the reachable range in density is approximately halved.

Besides smaller cancellations between competing terms, a smaller volume more easily admits rare configurations that are more
likely to be relevant at higher density. In contrast, in a larger volume high density physics appears
only as local fluctuations. One way to enrich the sampled configuration space
with data from the high density phase, is to introduce global constraints favoring
these. This can be in principle be done with density-of-states type algorithms, which have already
found applications in the context of QCD thermodynamics \cite{Fodor:2007vv,Ejiri:2008xt,Borsanyi:2021gqg,Lucini:2023irm}.

\begin{acknowledgments} 
We thank Philippe de Forcrand and Volodymyr Vovchenko for fruitful discussions
and for commenting the manuscript. The authors also acknowledge exchanges
with Ivan Horvath and Jan M. Pawlowski on related subjects.
This work is supported by the
MKW NRW under the funding code NW21-024-A.
Z. Fodor acknowledges funding from the DOE under the contract number DE-SC0025025.
This work was also supported by the
Hungarian National Research, Development and Innovation
Office, NKFIH Grant No. KKP126769.
This work was also supported by the NKFIH excellence
grant TKP2021{\textunderscore}NKTA{\textunderscore}64.
This work is also supported by the Hungarian National Research,
Development and Innovation
Office under Project No. FK 147164.
The authors gratefully acknowledge the Gauss Centre for
Supercomputing e.V. (\url{www.gauss-centre.eu}) for funding
this project by providing computing time on the GCS Supercomputer
Juwels-Booster at Juelich Supercomputer Centre.
We acknowledge the EuroHPC Joint Undertaking for awarding this project access to the EuroHPC supercomputer LUMI, hosted by CSC (Finland) and the LUMI consortium through a EuroHPC Extreme Access call.
An award of computer time was provided by the INCITE program. 
This research used resources of the Argonne Leadership Computing Facility, which
is a DOE Office of Science User Facility supported under Contract DE-AC02-06CH11357.
\end{acknowledgments} 

\appendix
\section{Example: non-interacting hadron gas\label{app:skellam}}

In the low temperature phase QCD can be described with very good
approximation as a non-interacting hadron gas. Lattice studies
have confirmed the concept, that the fugacity series
\begin{eqnarray}
\chi_1(T,V,\mu) = \sum_{k=1}^{\infty} b_k(T) \sinh\left(k\frac{\mu}{T}\right)
\end{eqnarray}
converges fast, and already the second coefficient $b_2(T)$ 
can hardly be distinguished from zero below $T<T_c$
\cite{Vovchenko:2017xad,Huovinen:2017ogf,Bellwied:2021nrt}.
Thus, the simple ``cosh model''
\begin{eqnarray}
\hat p_{GC} \equiv \frac{1}{VT^3} \log Z_{GC} 
= \hat p_M(T) + \hat p_B(T) \cosh\left(\frac{\mu}{T}\right)
\label{eq:cosh}
\end{eqnarray}
makes a reasonable choice to illustrate the canonical formalism.
Here $\hat p_M$ and $\hat p_B\equiv b_1$ stand for the mesonic and baryonic
contributions, respectively, assuming the validity of the Boltzmann
approximation in the Hadron resonance gas picture. We will use
the ``hat'' notation to indicate a normalization to temperature:
\begin{eqnarray}
\hat p = \frac{p}{T^4}\,,\quad\hat\mu = \frac{\mu}{T}
\end{eqnarray}
It is well known, that the canonical partition function of the
cosh model (\ref{eq:cosh}) is given by the Skellam distribution.
\begin{eqnarray}
Z_C(T,V,N) &=& \int_0^{2\pi} \frac{d\varphi}{2\pi}
e^{-i\varphi N + VT^3 \hat p_M + VT^3 \hat p_B \cos(\varphi)}\nonumber\\
&=& e^{VT^3\hat p_M} \, I_N\left( VT^3 \hat p_B\right)\,
\end{eqnarray}
where $I_N(z)$ is the modified Bessel function of first kind.
The temperature dependence of $Z_C$ enters through the coefficients
$\hat p^M(T)$ and $\hat p^B(T)$.
This matches the general form of the Skellam distribution
\begin{eqnarray}
e^{-(\lambda_1+\lambda_2)}\left(\frac{\lambda_1}{\lambda_2}\right)^{N/2} I_N(2\sqrt{\lambda_1\lambda_2})
\label{eq:skellampmf}
\end{eqnarray}
with $\lambda_1=\lambda_2=\frac{1}{2} (LT)^3 \hat p_B$.
Mesons play no role.

The grand canonical relations, which are fully inherited by $F_{LT}$, such as
\begin{eqnarray}
N &=& VT^3 \hat p_B \sinh \hat\mu
\end{eqnarray}
are not exactly satisfied if we replace $\mu$ with $\mu^f$ or $\mu^b$.
Having an exact form for $Z_C(T,V,N)$ we are in the position to quantify
these canonical volume effects.

Let us start with the backward definition of the chemical potential
\begin{eqnarray}
\hat \mu^b(T) &=& F(T,V,N)-F(T,V,N-1) \nonumber\\
&=&
-\log\left(\frac{I_N(VT^3\hat p_B)}{I_{N-1}(VT^3\hat p_B)}\right)
\label{eq:mubderiv1}
\end{eqnarray}
We will keep the density $\hat n=N/(VT^3)$ fixed, which is set by
the chemical potential $\hat \mu_B$,
\begin{eqnarray}
\hat n = \hat p_B \sinh(\hat \mu_B) = \frac{\hat p_B}{c}\,,\qquad
c=\frac{1}{\sinh(\hat\mu_B)}\,.
\end{eqnarray}
Through the shorthand $c$ we can simply write
\begin{eqnarray}
\hat \mu^b &=& \log\left(\frac{I_{N-1}(Nc)}{I_{N}(Nc)}\right)\,
\label{eq:mubderiv2}
\end{eqnarray}
where we can now take the $N\to \infty$ thermodynamic limit.
The asymptotic formula for the Bessel functions with simultaneously
growing index and argument reads (9.7.7 of Ref.~\cite{AS:1972})
\begin{eqnarray}
I_{N}(z)&=&
\frac{1}{\sqrt{2\pi z}(1+\frac{N^{2}}{z^{2}})^{1/4}}
\exp\Bigg(-N\,\asinh\bigg(\frac{N}{z}\bigg)\nonumber\\
&&+z\sqrt{1+\frac{N^{2}}{z^{2}}}\Bigg)
\Bigg(1+\mathcal{O}\big(\frac{1}{z\sqrt{1+\frac{N^{2}}{z^{2}}}}\big)\Bigg)
\label{eq:INasymp}
\end{eqnarray}
Inserting $z=Nc$ we have
\begin{eqnarray}
I_N(Nc)&=&
\frac{1}{\sqrt{2\pi N}(1+c^{2})^{1/4}}\exp\Bigg(-N\mu_{B}/T\nonumber\\
&&+N\sqrt{1+c^{2}}\Bigg)\Bigg(1+\mathcal{O}\big(N^{-1}\big)\Bigg)\,.
\end{eqnarray}
The same asymptotic form was used in \cite{Bzdak:2025rhp} to derive baryon/antibaryon cumulant in canonical HRG.
With this asymptotic form we can calculate $F$ near the thermodynamic limit:
\begin{eqnarray}
    \frac{F\left(T,V=N \frac{c}{\hat p_B T^3},N\right)}{VT^4}
    &=& - \hat p_M(T) - \frac{1}{VT^3} \log I_N(Nc) \nonumber\\
    &=&-\hat p_M(T) + \frac{\hat p_B}{Nc} \log \sqrt{2\pi N}(1+c^2)^{1/4}
    + \hat p_B\frac{ \hat \mu_B -\sqrt{1+c^2}}{c} + \frac{\hat p_B}{Nc} \mathcal{O}(N^{-1})\nonumber\\
    &=& \frac{1}{VT^4} \left(\Omega + \mu N\right) 
    -\frac{\hat p_B}{2Nc}\log\frac{1}{2\pi N} - \frac{\hat p_B}{2 N} \sinh(\hat\mu_B)\log\tanh{\hat \mu_B} + \mathcal{O}(N^{-2})
\end{eqnarray}
The corresponding generic formula was given in Eq.~(\ref{eq:FLTasymp}).
The $N\to\infty$ limit has to be understood here at constant density,
that is $c=1/\sinh\hat\mu_B$ fixed. 

\begin{figure*}
\centering
\includegraphics[width=0.30\textwidth]{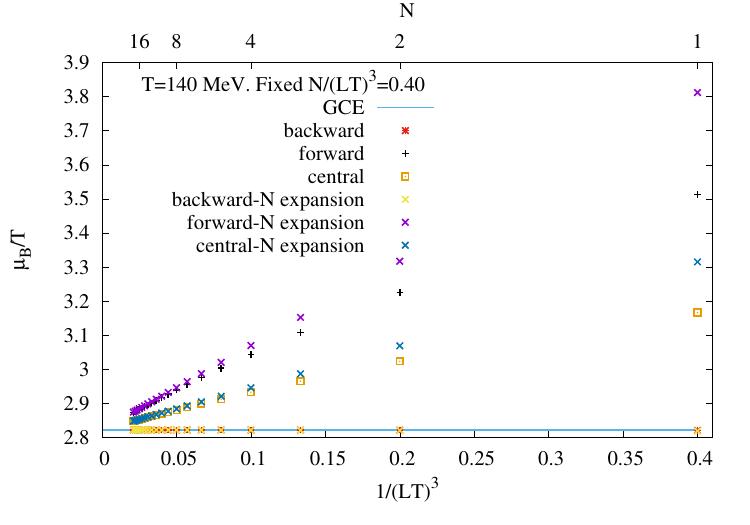}
\includegraphics[width=0.30\textwidth]{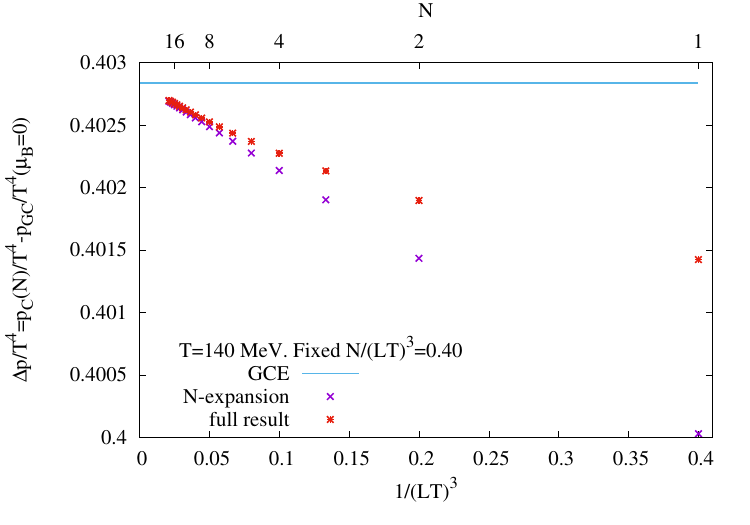}
\includegraphics[width=0.30\textwidth]{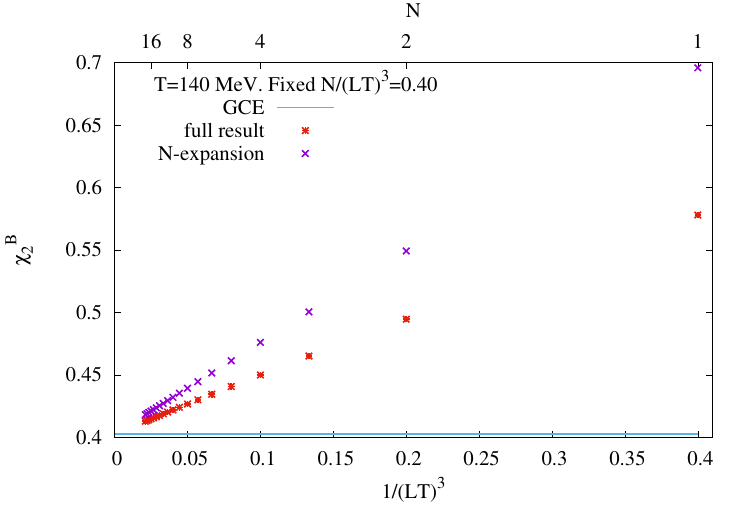}
\caption{\label{fig:skellammu}
Left:
The forward ($\mu^f$) and backward ($\mu^b$) chemical potentials in a finite
volume in a concrete example. $\hat p_B$ is taken from lattice
at $T=140$~MeV. The volume independent grand canonical value
$\hat\mu=\asinh(\hat n/\hat p_B)$ is shown as a line ``GCE''. We also show the
central chemical potential ($(\mu^f+\mu^b)/2$)
and evaluate the leading asymptotic formulas text to the exact results
in Eqs.~(\ref{eq:muf}-\ref{eq:mub}).
Center: the analogous result for the pressure at finite volume.
Right: the corresponding $\chi_2$ coefficient.
}
\end{figure*}

Let us now show that the observables follow the $1/N$ asymptotics,
as announced.
Applying the asymptotic form to $I_{N\pm1}(Nc)$ as well, simple algebra
yields
\begin{eqnarray}
\hat{\mu}^{f}(N)&=&\hat{\mu}+\frac{1}{2N}\frac{1}{\sqrt{1+c(\hat{\mu})^{2}}}+\frac{1}{2N(1+c(\hat{\mu})^{2})}+\mathcal{O}(\frac{1}{N^{2}})\label{eq:muhatf_Nc}\nonumber
\\
\hat{\mu}^{b}(N)&=&\hat{\mu}-\frac{1}{2N}\frac{1}{\sqrt{1+c(\hat{\mu})^{2}}}+\frac{1}{2N(1+c(\hat{\mu})^{2})}+\mathcal{O}(\frac{1}{N^{2}})\label{eq:muhatb_Nc}\nonumber\\
&&\label{eq:muhatfb}
\end{eqnarray}
For small $c$ (i.e. $\mu\gtrsim 1$) there is an approximate cancellation
between the two leading correction terms of $\hat\mu^b$. In contrast the contributions
amplify for $\hat\mu^f$. If we re-expand in $c$, this behavior is exposed as
\begin{eqnarray}
\hat{\mu}^{f}(N)&=&\hat{\mu}+\frac{1}{N}-\frac{3c^{2}}{4N}+\mathcal{O}(N^{-2})+\mathcal{O}(c^{4})\label{eq:muhatf_exp_c}\\
\hat{\mu}^{b}(N)&=&\hat{\mu}-\frac{c^{2}}{4N}+\mathcal{O}(N^{-2})+\mathcal{O}(c^{4})\label{eq:muhatb_exp_c}
\end{eqnarray}
We plot the formulas with realistic parameters in Fig.~\ref{fig:skellammu}.

The pressure (see Eq.~(\ref{eq:pcanon})) is computed with the same principle
\begin{equation}
\frac{p_{C}(T,V,N)}{T^{4}}=\frac{p_{GC}(T,V,\mu)}{T^{4}}-\hat{p}_{B}
e^{\hat{\mu}}(\hat{\mu}-\hat{\mu}^{b})+\mathcal{O}(N^{-2})\,,
\end{equation}
here the leading $\sim 1/N$ terms are encoded in $\hat{\mu}^{b}$.

We can also extract $\chi_2=\hat p_B$ as
\begin{eqnarray}
\frac{1}{\chi_2^{\rm discrete}(N)} &\equiv& 
(VT^3)(\hat\mu^f(N)-\hat\mu^b(N))
\end{eqnarray}
For the leading asymptotics in $1/\chi_2$ we will need one more order in the $1/N$
expression of $\hat \mu^b$ and $\hat \mu^f$. A tedious computation then gives
\begin{eqnarray}
\frac{1}{\chi_2^{\rm discrete}(N)}=\frac{1}{\hat p_B\cosh(\hat\mu)(1-\frac{1}{2N\sqrt{1+c^{2}}}+\frac{5}{4}\frac{1}{(1+c^{2})^{3/2}}\frac{1}{N})}+O({N^{-2}})
\end{eqnarray}
recovering the grand canonical result in the $N\to\infty$ limit with an
$\mathcal{O}(1/N)$ correction. The actual numerical result on the finite $N$
observables, and their approach to the grand canonical limit we show in Fig.~\ref{fig:skellammu}.

\bibliographystyle{unsrt}
\bibliography{thermo}

@article{Endrodi:2026tqa,
	archiveprefix = {arXiv},
	author = {Endrodi, Gergely and Moore, Guy D. and Pieczynski, Adam and Sciarra, Alessandro},
	eprint = {2603.22046},
	month = {3},
	primaryclass = {hep-lat},
	title = {{Exact center symmetry and first-order phase transition in QCD with three degenerate dynamical quarks}},
	year = {2026}}

@article{Fischer:2026uni,
	archiveprefix = {arXiv},
	author = {Fischer, Christian S. and Pawlowski, Jan M.},
	eprint = {2603.11135},
	month = {3},
	primaryclass = {hep-ph},
	title = {{Phase structure and observables at high densities from first principles QCD}},
	year = {2026}}

@book{AS:1972,
	author = {Milton Abramowitz, Irene A. Stegun},
	publisher = {National Bureau of Standards},
	title = {Handbook of Mathematical Functions},
	year = {1972}}

@article{Barbour:1997bh,
	archiveprefix = {arXiv},
	author = {Barbour, Ian M. and Morrison, Susan E. and Klepfish, Elyakum G. and Kogut, John B. and Lombardo, Maria-Paola},
	doi = {10.1103/PhysRevD.56.7063},
	eprint = {hep-lat/9705038},
	journal = {Phys. Rev. D},
	pages = {7063--7072},
	title = {{The Critical points of strongly coupled lattice QCD at nonzero chemical potential}},
	volume = {56},
	year = {1997}}

@article{Kratochvila:2006jx,
	archiveprefix = {arXiv},
	author = {Kratochvila, Slavo and de Forcrand, Philippe},
	doi = {10.1103/PhysRevD.73.114512},
	eprint = {hep-lat/0602005},
	journal = {Phys. Rev. D},
	pages = {114512},
	reportnumber = {CERN-PH-TH-2006-007},
	title = {{QCD at zero baryon density and the Polyakov loop paradox}},
	volume = {73},
	year = {2006}}

@article{Adam:2025hpb,
	author = {Alexander Adam and Szabolcs Bors{\'a}nyi and Zoltan Fodor and Jana N. Guenther and Piyush Kumar and Paolo Parotto and Attila P{\'a}sztor and Chik Him Wong},
	eprint = {2507.13254},
	month = {07},
	title = {High-precision baryon number cumulants from lattice QCD in a finite box: cumulant ratios, Lee-Yang zeros and critical endpoint predictions},
	url = {https://arxiv.org/pdf/2507.13254.pdf},
	year = {2025}}

@article{Lucini:2023irm,
	archiveprefix = {arXiv},
	author = {Lucini, Biagio and Mason, David and Piai, Maurizio and Rinaldi, Enrico and Vadacchino, Davide},
	doi = {10.1103/PhysRevD.108.074517},
	eprint = {2305.07463},
	journal = {Phys. Rev. D},
	number = {7},
	pages = {074517},
	primaryclass = {hep-lat},
	reportnumber = {RIKEN-iTHEMS-Report-23 ET-0164A-23},
	title = {{First-order phase transitions in Yang-Mills theories and the density of state method}},
	volume = {108},
	year = {2023}}

@article{Bollweg:2022fqq,
	archiveprefix = {arXiv},
	author = {Bollweg, D. and Clarke, D. A. and Goswami, J. and Kaczmarek, O. and Karsch, F. and Mukherjee, Swagato and Petreczky, P. and Schmidt, C. and Sharma, Sipaz},
	collaboration = {HotQCD},
	doi = {10.1103/PhysRevD.108.014510},
	eprint = {2212.09043},
	journal = {Phys. Rev. D},
	number = {1},
	pages = {014510},
	primaryclass = {hep-lat},
	title = {{Equation of state and speed of sound of (2+1)-flavor QCD in strangeness-neutral matter at nonvanishing net baryon-number density}},
	volume = {108},
	year = {2023}}

@article{Vovchenko:2019pjl,
    author = "Vovchenko, Volodymyr and Stoecker, Horst",
    title = "{Thermal-FIST: A package for heavy-ion collisions and hadronic equation of state}",
    eprint = "1901.05249",
    archivePrefix = "arXiv",
    primaryClass = "nucl-th",
    doi = "10.1016/j.cpc.2019.06.024",
    journal = "Comput. Phys. Commun.",
    volume = "244",
    pages = "295--310",
    year = "2019"
}

@article{Borsanyi:2024xrx,
	archiveprefix = {arXiv},
	author = {Borsanyi, Szabolcs and Fodor, Zoltan and Guenther, Jana N. and Parotto, Paolo and Pasztor, Attila and Pirelli, Ludovica and Szabo, Kalman K. and Wong, Chik Him},
	doi = {10.1103/PhysRevD.110.114507},
	eprint = {2410.06216},
	journal = {Phys. Rev. D},
	number = {11},
	pages = {114507},
	primaryclass = {hep-lat},
	title = {{QCD deconfinement transition line up to \ensuremath{\mu}B=400\,\,MeV from finite volume lattice simulations}},
	volume = {110},
	year = {2024}}

@article{Borsanyi:2023tdp,
	archiveprefix = {arXiv},
	author = {Borsanyi, Szabolcs and Fodor, Zoltan and Giordano, Matteo and Guenther, Jana N. and Katz, Sandor D. and Pasztor, Attila and Wong, Chik Him},
	doi = {10.1103/PhysRevD.109.054509},
	eprint = {2308.06105},
	journal = {Phys. Rev. D},
	number = {5},
	pages = {054509},
	primaryclass = {hep-lat},
	title = {{Can rooted staggered fermions describe nonzero baryon density at low temperatures?}},
	volume = {109},
	year = {2024}}

@article{deForcrand:2006ec,
	archiveprefix = {arXiv},
	author = {de Forcrand, Philippe and Kratochvila, Slavo},
	doi = {10.1016/j.nuclphysbps.2006.01.007},
	editor = {Alexandrou, Constantia and Panagopoulos, Haralambos and Schierholz, Gerrit},
	eprint = {hep-lat/0602024},
	journal = {Nucl. Phys. B Proc. Suppl.},
	pages = {62--67},
	title = {{Finite density QCD with a canonical approach}},
	volume = {153},
	year = {2006}}

@article{Li:2010qf,
	archiveprefix = {arXiv},
	author = {Li, Anyi and Alexandru, Andrei and Liu, Keh-Fei and Meng, Xiangfei},
	doi = {10.1103/PhysRevD.82.054502},
	eprint = {1005.4158},
	journal = {Phys. Rev. D},
	pages = {054502},
	primaryclass = {hep-lat},
	reportnumber = {UK-10-04},
	title = {{Finite density phase transition of QCD with $N_f=4$ and $N_f=2$ using canonical ensemble method}},
	volume = {82},
	year = {2010}}

@article{Borsanyi:2023wno,
	archiveprefix = {arXiv},
	author = {Borsanyi, Szabolcs and Fodor, Zoltan and Guenther, Jana N. and Katz, Sandor D. and Parotto, Paolo and Pasztor, Attila and Pesznyak, David and Szabo, Kalman K. and Wong, Chik Him},
	doi = {10.1103/PhysRevD.110.L011501},
	eprint = {2312.07528},
	journal = {Phys. Rev. D},
	number = {1},
	pages = {L011501},
	primaryclass = {hep-lat},
	title = {{Continuum-extrapolated high-order baryon fluctuations}},
	volume = {110},
	year = {2024}}

@article{Gunkel:2021oya,
	archiveprefix = {arXiv},
	author = {Gunkel, Pascal J. and Fischer, Christian S.},
	doi = {10.1103/PhysRevD.104.054022},
	eprint = {2106.08356},
	journal = {Phys. Rev. D},
	number = {5},
	pages = {054022},
	primaryclass = {hep-ph},
	title = {{Locating the critical endpoint of QCD: Mesonic backcoupling effects}},
	volume = {104},
	year = {2021}}

@article{Son:2000xc,
	archiveprefix = {arXiv},
	author = {Son, D. T. and Stephanov, Misha A.},
	doi = {10.1103/PhysRevLett.86.592},
	eprint = {hep-ph/0005225},
	journal = {Phys. Rev. Lett.},
	pages = {592--595},
	title = {{QCD at finite isospin density}},
	volume = {86},
	year = {2001}}

@article{Bollweg:2022rps,
	archiveprefix = {arXiv},
	author = {Bollweg, D. and Goswami, J. and Kaczmarek, O. and Karsch, F. and Mukherjee, Swagato and Petreczky, P. and Schmidt, C. and Scior, P.},
	collaboration = {HotQCD},
	doi = {10.1103/PhysRevD.105.074511},
	eprint = {2202.09184},
	journal = {Phys. Rev. D},
	number = {7},
	pages = {074511},
	primaryclass = {hep-lat},
	title = {{Taylor expansions and Pad\'e approximants for cumulants of conserved charge fluctuations at nonvanishing chemical potentials}},
	volume = {105},
	year = {2022}}

@article{Borsanyi:2022soo,
	archiveprefix = {arXiv},
	author = {Borsanyi, Szabolcs and Fodor, Zoltan and Giordano, Matteo and Guenther, Jana N. and Katz, Sandor D. and Pasztor, Attila and Wong, Chik Him},
	doi = {10.1103/PhysRevD.107.L091503},
	eprint = {2208.05398},
	journal = {Phys. Rev. D},
	number = {9},
	pages = {L091503},
	primaryclass = {hep-lat},
	title = {{Equation of state of a hot-and-dense quark gluon plasma: Lattice simulations at real \ensuremath{\mu}B vs extrapolations}},
	volume = {107},
	year = {2023}}

@article{Gao:2020fbl,
	archiveprefix = {arXiv},
	author = {Gao, Fei and Pawlowski, Jan M.},
	doi = {10.1016/j.physletb.2021.136584},
	eprint = {2010.13705},
	journal = {Phys. Lett. B},
	pages = {136584},
	primaryclass = {hep-ph},
	title = {{Chiral phase structure and critical end point in QCD}},
	volume = {820},
	year = {2021}}

@article{Borsanyi:2021hbk,
	archiveprefix = {arXiv},
	author = {Borsanyi, Szabolcs and Fodor, Zoltan and Giordano, Matteo and Katz, Sandor D. and Nogradi, Daniel and Pasztor, Attila and Wong, Chik Him},
	doi = {10.1103/PhysRevD.105.L051506},
	eprint = {2108.09213},
	journal = {Phys. Rev. D},
	number = {5},
	pages = {L051506},
	primaryclass = {hep-lat},
	title = {{Lattice simulations of the QCD chiral transition at real baryon density}},
	volume = {105},
	year = {2022}}

@article{Bellwied:2021nrt,
	archiveprefix = {arXiv},
	author = {Bellwied, Rene and Borsanyi, Szabolcs and Fodor, Zoltan and Guenther, Jana N. and Katz, Sandor D. and Parotto, Paolo and Pasztor, Attila and Pesznyak, David and Ratti, Claudia and Szabo, Kalman K.},
	doi = {10.1103/PhysRevD.104.094508},
	eprint = {2102.06625},
	journal = {Phys. Rev. D},
	number = {9},
	pages = {094508},
	primaryclass = {hep-lat},
	title = {{Corrections to the hadron resonance gas from lattice QCD and their effect on fluctuation-ratios at finite density}},
	volume = {104},
	year = {2021}}

@article{Bollweg:2021vqf,
	archiveprefix = {arXiv},
	author = {Bollweg, D. and Goswami, J. and Kaczmarek, O. and Karsch, F. and Mukherjee, Swagato and Petreczky, P. and Schmidt, C. and Scior, P.},
	collaboration = {HotQCD},
	doi = {10.1103/PhysRevD.104.074512},
	eprint = {2107.10011},
	journal = {Phys. Rev. D},
	number = {7},
	primaryclass = {hep-lat},
	title = {{Second order cumulants of conserved charge fluctuations revisited: Vanishing chemical potentials}},
	volume = {104},
	year = {2021}}

@article{Borsanyi:2021sxv,
	archiveprefix = {arXiv},
	author = {Bors\'anyi, S. and Fodor, Z. and Guenther, J. N. and Kara, R. and Katz, S. D. and Parotto, P. and P\'asztor, A. and Ratti, C. and Szab\'o, K. K.},
	doi = {10.1103/PhysRevLett.126.232001},
	eprint = {2102.06660},
	journal = {Phys. Rev. Lett.},
	number = {23},
	pages = {232001},
	primaryclass = {hep-lat},
	title = {{Lattice QCD equation of state at finite chemical potential from an alternative expansion scheme}},
	volume = {126},
	year = {2021}}

@article{Borsanyi:2021gqg,
	archiveprefix = {arXiv},
	author = {Borsanyi, Szablocs and Sexty, D\'enes},
	doi = {10.1016/j.physletb.2021.136148},
	eprint = {2101.03383},
	journal = {Phys. Lett. B},
	pages = {136148},
	primaryclass = {hep-lat},
	title = {{Topological susceptibility of pure gauge theory using Density of States}},
	volume = {815},
	year = {2021}}

@article{Li:2011ee,
	archiveprefix = {arXiv},
	author = {Li, Anyi and Alexandru, Andrei and Liu, Keh-Fei},
	doi = {10.1103/PhysRevD.84.071503},
	eprint = {1103.3045},
	journal = {Phys. Rev. D},
	pages = {071503},
	primaryclass = {hep-ph},
	title = {{Critical point of $N_f = 3$ QCD from lattice simulations in the canonical ensemble}},
	volume = {84},
	year = {2011}}

@article{Fu:2019hdw,
	archiveprefix = {arXiv},
	author = {Fu, Wei-jie and Pawlowski, Jan M. and Rennecke, Fabian},
	doi = {10.1103/PhysRevD.101.054032},
	eprint = {1909.02991},
	journal = {Phys. Rev. D},
	number = {5},
	pages = {054032},
	primaryclass = {hep-ph},
	title = {{QCD phase structure at finite temperature and density}},
	volume = {101},
	year = {2020}}

@article{Giordano:2019gev,
	archiveprefix = {arXiv},
	author = {Giordano, Matteo and Kapas, Kornel and Katz, Sandor D. and Nogradi, Daniel and Pasztor, Attila},
	doi = {10.1103/PhysRevD.101.074511},
	eprint = {1911.00043},
	journal = {Phys. Rev. D},
	number = {7},
	pages = {074511},
	primaryclass = {hep-lat},
	title = {{Radius of convergence in lattice QCD at finite $\mu_B$ with rooted staggered fermions}},
	volume = {101},
	year = {2020}}

@article{Giordano:2020roi,
	archiveprefix = {arXiv},
	author = {Giordano, Matteo and Kapas, Kornel and Katz, Sandor D. and Nogradi, Daniel and Pasztor, Attila},
	doi = {10.1007/JHEP05(2020)088},
	eprint = {2004.10800},
	journal = {JHEP},
	pages = {088},
	primaryclass = {hep-lat},
	title = {{New approach to lattice QCD at finite density; results for the critical end point on coarse lattices}},
	volume = {05},
	year = {2020}}

@article{Borsanyi:2020fev,
	archiveprefix = {arXiv},
	author = {Borsanyi, Szabolcs and Fodor, Zoltan and Guenther, Jana N. and Kara, Ruben and Katz, Sandor D. and Parotto, Paolo and Pasztor, Attila and Ratti, Claudia and Szabo, Kalman K.},
	eprint = {2002.02821},
	journal = {Phys. Rev. Lett.},
	pages = {052001},
	primaryclass = {hep-lat},
	slaccitation = {%%CITATION = ARXIV:2002.02821;%%},
	title = {{The QCD crossover at finite chemical potential from lattice simulations}},
	volume = {125},
	year = {2020}}

@article{Fischer:2018sdj,
	archiveprefix = {arXiv},
	author = {Fischer, Christian S.},
	doi = {10.1016/j.ppnp.2019.01.002},
	eprint = {1810.12938},
	journal = {Prog. Part. Nucl. Phys.},
	pages = {1-60},
	primaryclass = {hep-ph},
	slaccitation = {%%CITATION = ARXIV:1810.12938;%%},
	title = {{QCD at finite temperature and chemical potential from DysonSchwinger equations}},
	volume = {105},
	year = {2019}}

@article{Bazavov:2018mes,
	archiveprefix = {arXiv},
	author = {Bazavov, A. and others},
	eprint = {1812.08235},
	journal = {Physics Letters B},
	month = {Aug},
	pages = {15--21},
	primaryclass = {hep-lat},
	slaccitation = {%%CITATION = ARXIV:1812.08235;%%},
	title = {{Chiral crossover in QCD at zero and non-zero chemical potentials}},
	volume = {795},
	year = {2019}}

@article{Borsanyi:2018grb,
	archiveprefix = {arXiv},
	author = {Borsanyi, Szabolcs and Fodor, Zoltan and Guenther, Jana N. and Katz, Sandor K. and Szabo, Kalman K. and Pasztor, Attila and Portillo, Israel and Ratti, Claudia},
	doi = {10.1007/JHEP10(2018)205},
	eprint = {1805.04445},
	journal = {JHEP},
	pages = {205},
	primaryclass = {hep-lat},
	slaccitation = {%%CITATION = ARXIV:1805.04445;%%},
	title = {{Higher order fluctuations and correlations of conserved charges from lattice QCD}},
	volume = {10},
	year = {2018}}

@article{Huovinen:2017ogf,
	archiveprefix = {arXiv},
	author = {Huovinen, Pasi and Petreczky, Peter},
	doi = {10.1016/j.physletb.2017.12.001},
	eprint = {1708.00879},
	journal = {Phys. Lett.},
	pages = {125-130},
	primaryclass = {hep-ph},
	slaccitation = {%%CITATION = ARXIV:1708.00879;%%},
	title = {{Hadron Resonance Gas with Repulsive Interactions and Fluctuations of Conserved Charges}},
	volume = {B777},
	year = {2018}}

@article{Barbour:1997ej,
	archiveprefix = {arXiv},
	author = {Barbour, Ian M. and Morrison, Susan E. and Klepfish, Elyakum G. and Kogut, John B. and Lombardo, Maria-Paola},
	booktitle = {{Lattice QCD on parallel computers. Proceedings, International Workshop, Tsukuba, Japan, March 10-15, 1997}},
	doi = {10.1016/S0920-5632(97)00484-2},
	eprint = {hep-lat/9705042},
	journal = {Nucl. Phys. Proc. Suppl.},
	note = {[,220(1997)]},
	pages = {220-234},
	primaryclass = {hep-lat},
	slaccitation = {%%CITATION = HEP-LAT/9705042;%%},
	title = {{Results on finite density QCD}},
	volume = {60A},
	year = {1998}}

@article{Alford:1998sd,
	archiveprefix = {arXiv},
	author = {Alford, Mark G. and Kapustin, Anton and Wilczek, Frank},
	doi = {10.1103/PhysRevD.59.054502},
	eprint = {hep-lat/9807039},
	journal = {Phys. Rev.},
	pages = {054502},
	primaryclass = {hep-lat},
	reportnumber = {IASSNS-HEP-98-67},
	slaccitation = {%%CITATION = HEP-LAT/9807039;%%},
	title = {{Imaginary chemical potential and finite fermion density on the lattice}},
	volume = {D59},
	year = {1999}}

@article{Vovchenko:2017xad,
	archiveprefix = {arXiv},
	author = {Vovchenko, Volodymyr and Pasztor, Attila and Fodor, Zoltan and Katz, Sandor D. and Stoecker, Horst},
	doi = {10.1016/j.physletb.2017.10.042},
	eprint = {1708.02852},
	journal = {Phys. Lett.},
	pages = {71-78},
	primaryclass = {hep-ph},
	slaccitation = {%%CITATION = ARXIV:1708.02852;%%},
	title = {{Repulsive baryonic interactions and lattice QCD observables at imaginary chemical potential}},
	volume = {B775},
	year = {2017}}

@article{Hasenfratz:1991ax,
	author = {Hasenfratz, A. and Toussaint, D.},
	doi = {10.1016/0550-3213(92)90247-9},
	journal = {Nucl. Phys.},
	pages = {539-549},
	reportnumber = {AZPH-TH-91-21},
	slaccitation = {%%CITATION = NUPHA,B371,539;%%},
	title = {{Canonical ensembles and nonzero density quantum chromodynamics}},
	volume = {B371},
	year = {1992}}

@article{DElia:2016jqh,
	archiveprefix = {arXiv},
	author = {D'Elia, Massimo and Gagliardi, Giuseppe and Sanfilippo, Francesco},
	doi = {10.1103/PhysRevD.95.094503},
	eprint = {1611.08285},
	journal = {Phys. Rev.},
	number = {9},
	pages = {094503},
	primaryclass = {hep-lat},
	slaccitation = {%%CITATION = ARXIV:1611.08285;%%},
	title = {{Higher order quark number fluctuations via imaginary chemical potentials in $N_f=2+1$ QCD}},
	volume = {D95},
	year = {2017}}

@article{Fukuda:2015mva,
	archiveprefix = {arXiv},
	author = {Fukuda, Ryutaro and Nakamura, Atsushi and Oka, Shotaro},
	doi = {10.1103/PhysRevD.93.094508},
	eprint = {1504.06351},
	journal = {Phys. Rev.},
	number = {9},
	pages = {094508},
	primaryclass = {hep-lat},
	slaccitation = {%%CITATION = ARXIV:1504.06351;%%},
	title = {{Canonical approach to finite density QCD with multiple precision computation}},
	volume = {D93},
	year = {2016}}

@article{Bellwied:2015rza,
	archiveprefix = {arXiv},
	author = {Bellwied, R. and Borsanyi, S. and Fodor, Z. and G{\"u}nther, J. and Katz, S. D. and Ratti, C. and Szabo, K. K.},
	doi = {10.1016/j.physletb.2015.11.011},
	eprint = {1507.07510},
	journal = {Phys. Lett.},
	pages = {559-564},
	primaryclass = {hep-lat},
	slaccitation = {%%CITATION = ARXIV:1507.07510;%%},
	title = {{The QCD phase diagram from analytic continuation}},
	volume = {B751},
	year = {2015}}

@article{Bellwied:2015lba,
	archiveprefix = {arXiv},
	author = {Bellwied, R. and Borsanyi, S. and Fodor, Z. and Katz, S. D. and Pasztor, A. and Ratti, C. and Szabo, K. K.},
	doi = {10.1103/PhysRevD.92.114505},
	eprint = {1507.04627},
	journal = {Phys. Rev.},
	number = {11},
	pages = {114505},
	primaryclass = {hep-lat},
	slaccitation = {%%CITATION = ARXIV:1507.04627;%%},
	title = {{Fluctuations and correlations in high temperature QCD}},
	volume = {D92},
	year = {2015}}

@article{Ding:2015fca,
	archiveprefix = {arXiv},
	author = {Ding, H. -T. and Mukherjee, Swagato and Ohno, H. and Petreczky, P. and Schadler, H. -P.},
	doi = {10.1103/PhysRevD.92.074043},
	eprint = {1507.06637},
	journal = {Phys. Rev.},
	number = {7},
	pages = {074043},
	primaryclass = {hep-lat},
	slaccitation = {%%CITATION = ARXIV:1507.06637;%%},
	title = {{Diagonal and off-diagonal quark number susceptibilities at high temperatures}},
	volume = {D92},
	year = {2015}}

@article{Bonati:2015bha,
	archiveprefix = {arXiv},
	author = {Bonati, Claudio and D'Elia, Massimo and Mariti, Marco and Mesiti, Michele and Negro, Francesco and Sanfilippo, Francesco},
	doi = {10.1103/PhysRevD.92.054503},
	eprint = {1507.03571},
	journal = {Phys. Rev.},
	number = {5},
	pages = {054503},
	primaryclass = {hep-lat},
	reportnumber = {IFUP-TH-2015-07},
	slaccitation = {%%CITATION = ARXIV:1507.03571;%%},
	title = {{Curvature of the chiral pseudocritical line in QCD: Continuum extrapolated results}},
	volume = {D92},
	year = {2015}}

@article{Alexandru:2005ix,
	archiveprefix = {arXiv},
	author = {Alexandru, Andrei and Faber, Manfried and Horvath, Ivan and Liu, Keh-Fei},
	doi = {10.1103/PhysRevD.72.114513},
	eprint = {hep-lat/0507020},
	journal = {Phys. Rev.},
	pages = {114513},
	primaryclass = {hep-lat},
	reportnumber = {UK-05-07},
	slaccitation = {%%CITATION = HEP-LAT/0507020;%%},
	title = {{Lattice QCD at finite density via a new canonical approach}},
	volume = {D72},
	year = {2005}}

@article{Stephanov:1998dy,
	archiveprefix = {arXiv},
	author = {Stephanov, Misha A. and Rajagopal, K. and Shuryak, Edward V.},
	doi = {10.1103/PhysRevLett.81.4816},
	eprint = {hep-ph/9806219},
	journal = {Phys. Rev. Lett.},
	pages = {4816-4819},
	primaryclass = {hep-ph},
	reportnumber = {ITP-SB-98-39, MIT-CTP-2748, SUNY-NTG-98-17},
	slaccitation = {%%CITATION = HEP-PH/9806219;%%},
	title = {{Signatures of the tricritical point in QCD}},
	volume = {81},
	year = {1998}}

@article{Golterman:2006rw,
	archiveprefix = {arXiv},
	author = {Golterman, Maarten and Shamir, Yigal and Svetitsky, Benjamin},
	doi = {10.1103/PhysRevD.74.071501},
	eprint = {hep-lat/0602026},
	journal = {Phys. Rev.},
	pages = {071501},
	primaryclass = {hep-lat},
	reportnumber = {SINP-TNP-06-01},
	slaccitation = {%%CITATION = HEP-LAT/0602026;%%},
	title = {{Breakdown of staggered fermions at nonzero chemical potential}},
	volume = {D74},
	year = {2006}}

@article{Ejiri:2008xt,
	archiveprefix = {arXiv},
	author = {Ejiri, Shinji},
	doi = {10.1103/PhysRevD.78.074507},
	eprint = {0804.3227},
	journal = {Phys.Rev.},
	pages = {074507},
	primaryclass = {hep-lat},
	reportnumber = {BNL-NT-08-11},
	slaccitation = {%%CITATION = ARXIV:0804.3227;%%},
	title = {{Canonical partition function and finite density phase transition in lattice QCD}},
	volume = {D78},
	year = {2008}}

@article{Roberge:1986mm,
	author = {Roberge, Andre and Weiss, Nathan},
	doi = {10.1016/0550-3213(86)90582-1},
	journal = {Nucl.Phys.},
	pages = {734},
	reportnumber = {PRINT-86-0360 (BRITISH-COLUMBIA)},
	slaccitation = {%%CITATION = NUPHA,B275,734;%%},
	title = {{Gauge Theories With Imaginary Chemical Potential and the Phases of {QCD}}},
	volume = {B275},
	year = {1986}}

@article{Danzer:2012vw,
	archiveprefix = {arXiv},
	author = {Danzer, Julia and Gattringer, Christof},
	doi = {10.1103/PhysRevD.86.014502},
	eprint = {1204.1020},
	journal = {Phys.Rev.},
	pages = {014502},
	primaryclass = {hep-lat},
	reportnumber = {INT-PUB-12-016},
	slaccitation = {%%CITATION = ARXIV:1204.1020;%%},
	title = {{Properties of canonical determinants and a test of fugacity expansion for finite density lattice QCD with Wilson fermions}},
	volume = {D86},
	year = {2012}}

@article{Alexandru:2014hga,
	archiveprefix = {arXiv},
	author = {Alexandru, Andrei and Gattringer, C. and Schadler, H. -P. and Splittorff, K. and Verbaarschot, J.J.M.},
	doi = {10.1103/PhysRevD.91.074501},
	eprint = {1411.4143},
	journal = {Phys.Rev.},
	number = {7},
	pages = {074501},
	primaryclass = {hep-lat},
	slaccitation = {%%CITATION = ARXIV:1411.4143;%%},
	title = {{Distribution of Canonical Determinants in QCD}},
	volume = {D91},
	year = {2015}}

@article{Gavai:1989ce,
	author = {Gavai, R.V. and Potvin, J. and Sanielevici, S.},
	doi = {10.1103/PhysRevD.40.2743},
	journal = {Phys.Rev.},
	pages = {2743},
	reportnumber = {BUHEP-89-4, CERN-TH-5283-89, FSU-SCRI-89-18},
	slaccitation = {%%CITATION = PHRVA,D40,2743;%%},
	title = {{Quark Number Susceptibility in Quenched Quantum Chromodynamics}},
	volume = {D40},
	year = {1989}}

@article{Allton:2002zi,
	archiveprefix = {arXiv},
	author = {Allton, C.R. and Ejiri, S. and Hands, S.J. and Kaczmarek, O. and Karsch, F. and others},
	doi = {10.1103/PhysRevD.66.074507},
	eprint = {hep-lat/0204010},
	journal = {Phys.Rev.},
	pages = {074507},
	primaryclass = {hep-lat},
	reportnumber = {SWAT-02-335, NSF-ITP-02-26, BI-TP-2002-06},
	slaccitation = {%%CITATION = HEP-LAT/0204010;%%},
	title = {{The QCD thermal phase transition in the presence of a small chemical potential}},
	volume = {D66},
	year = {2002}}

@article{Hasenfratz:1983ba,
	author = {Hasenfratz, P. and Karsch, F.},
	doi = {10.1016/0370-2693(83)91290-X},
	journal = {Phys.Lett.},
	pages = {308},
	reportnumber = {CERN-TH-3530},
	slaccitation = {%%CITATION = PHLTA,B125,308;%%},
	title = {{Chemical Potential on the Lattice}},
	volume = {B125},
	year = {1983}}

@article{Borsanyi:2011sw,
	archiveprefix = {arXiv},
	author = {Borsanyi, Szabolcs and Fodor, Zoltan and Katz, Sandor D. and Krieg, Stefan and Ratti, Claudia and others},
	doi = {10.1007/JHEP01(2012)138},
	eprint = {1112.4416},
	journal = {JHEP},
	pages = {138},
	primaryclass = {hep-lat},
	slaccitation = {%%CITATION = ARXIV:1112.4416;%%},
	title = {{Fluctuations of conserved charges at finite temperature from lattice QCD}},
	volume = {1201},
	year = {2012}}

@article{Fodor:2007vv,
	archiveprefix = {arXiv},
	author = {Fodor, Zoltan and Katz, Sandor D. and Schmidt, Christian},
	doi = {10.1088/1126-6708/2007/03/121},
	eprint = {hep-lat/0701022},
	journal = {JHEP},
	pages = {121},
	primaryclass = {hep-lat},
	reportnumber = {WUP-0702, ITP-BUDAPEST-630, BNL-NT-07-8},
	slaccitation = {%%CITATION = HEP-LAT/0701022;%%},
	title = {{The Density of states method at non-zero chemical potential}},
	volume = {0703},
	year = {2007}}

@article{Aoki:2009sc,
	archiveprefix = {arXiv},
	author = {Aoki, Y. and Borsanyi, Szabolcs and Durr, Stephan and Fodor, Zoltan and Katz, Sandor D. and others},
	doi = {10.1088/1126-6708/2009/06/088},
	eprint = {0903.4155},
	journal = {JHEP},
	pages = {088},
	primaryclass = {hep-lat},
	reportnumber = {WUB-09-01, ITP-BUDAPEST-644, RBRC-782},
	slaccitation = {%%CITATION = ARXIV:0903.4155;%%},
	title = {{The QCD transition temperature: results with physical masses in the continuum limit II.}},
	volume = {0906},
	year = {2009}}

@article{Borsanyi:2010bp,
	archiveprefix = {arXiv},
	author = {Borsanyi, Szabolcs and others},
	collaboration = {Wuppertal-Budapest Collaboration},
	doi = {10.1007/JHEP09(2010)073},
	eprint = {1005.3508},
	journal = {JHEP},
	pages = {073},
	primaryclass = {hep-lat},
	reportnumber = {WUB-10-11, MIT-CTP-4152},
	slaccitation = {%%CITATION = ARXIV:1005.3508;%%},
	title = {{Is there still any $T_c$ mystery in lattice QCD? Results with physical masses in the continuum limit III}},
	volume = {1009},
	year = {2010}}

@article{Fodor:2001au,
	archiveprefix = {arXiv},
	author = {Fodor, Z. and Katz, S.D.},
	doi = {10.1016/S0370-2693(02)01583-6},
	eprint = {hep-lat/0104001},
	journal = {Phys.Lett.},
	pages = {87-92},
	primaryclass = {hep-lat},
	reportnumber = {DESY-01-044, ITP-BUDAPEST-566},
	slaccitation = {%%CITATION = HEP-LAT/0104001;%%},
	title = {{A New method to study lattice QCD at finite temperature and chemical potential}},
	volume = {B534},
	year = {2002}}

@article{Aoki:2006br,
	archiveprefix = {arXiv},
	author = {Aoki, Y. and Fodor, Z. and Katz, S.D. and Szabo, K.K.},
	doi = {10.1016/j.physletb.2006.10.021},
	eprint = {hep-lat/0609068},
	journal = {Phys.Lett.},
	pages = {46-54},
	primaryclass = {hep-lat},
	slaccitation = {%%CITATION = HEP-LAT/0609068;%%},
	title = {{The QCD transition temperature: Results with physical masses in the continuum limit}},
	volume = {B643},
	year = {2006}}

@article{Aoki:2006we,
	archiveprefix = {arXiv},
	author = {Aoki, Y. and Endrodi, G. and Fodor, Z. and Katz, S.D. and Szabo, K.K.},
	doi = {10.1038/nature05120},
	eprint = {hep-lat/0611014},
	journal = {Nature},
	pages = {675-678},
	primaryclass = {hep-lat},
	slaccitation = {%%CITATION = HEP-LAT/0611014;%%},
	title = {{The Order of the quantum chromodynamics transition predicted by the standard model of particle physics}},
	volume = {443},
	year = {2006}}

@article{Fodor:2001pe,
	archiveprefix = {arXiv},
	author = {Fodor, Z. and Katz, S.D.},
	doi = {10.1088/1126-6708/2002/03/014},
	eprint = {hep-lat/0106002},
	journal = {JHEP},
	pages = {014},
	primaryclass = {hep-lat},
	reportnumber = {ITP-BUDAPEST-568, DESY-01-057},
	slaccitation = {%%CITATION = HEP-LAT/0106002;%%},
	title = {{Lattice determination of the critical point of QCD at finite T and mu}},
	volume = {0203},
	year = {2002}}

@article{Yaffe:1982qf,
	author = {Yaffe, L.G. and Svetitsky, B.},
	doi = {10.1103/PhysRevD.26.963},
	journal = {Phys.Rev.},
	pages = {963},
	reportnumber = {CALT-68-913},
	slaccitation = {%%CITATION = PHRVA,D26,963;%%},
	title = {{First Order Phase Transition in the SU(3) Gauge Theory at Finite Temperature}},
	volume = {D26},
	year = {1982}}

@article{Bazavov:2011nk,
	archiveprefix = {arXiv},
	author = {Bazavov, A. and Bhattacharya, T. and Cheng, M. and DeTar, C. and Ding, H.T. and others},
	doi = {10.1103/PhysRevD.85.054503},
	eprint = {1111.1710},
	journal = {Phys.Rev.},
	pages = {054503},
	primaryclass = {hep-lat},
	slaccitation = {%%CITATION = ARXIV:1111.1710;%%},
	title = {{The chiral and deconfinement aspects of the QCD transition}},
	volume = {D85},
	year = {2012}}

@article{Bazavov:2012jq,
	archiveprefix = {arXiv},
	author = {Bazavov, A. and others},
	collaboration = {HotQCD Collaboration},
	doi = {10.1103/PhysRevD.86.034509},
	eprint = {1203.0784},
	journal = {Phys.Rev.},
	pages = {034509},
	primaryclass = {hep-lat},
	slaccitation = {%%CITATION = ARXIV:1203.0784;%%},
	title = {{Fluctuations and Correlations of net baryon number, electric charge, and strangeness: A comparison of lattice QCD results with the hadron resonance gas model}},
	volume = {D86},
	year = {2012}}

@article{BraunMunzinger:2003zz,
	archiveprefix = {arXiv},
	author = {Braun-Munzinger, P. and Stachel, J. and Wetterich, Christof},
	doi = {10.1016/j.physletb.2004.05.081},
	eprint = {nucl-th/0311005},
	journal = {Phys.Lett.},
	pages = {61-69},
	primaryclass = {nucl-th},
	reportnumber = {GSI-PREPRINT-2003-34},
	slaccitation = {%%CITATION = NUCL-TH/0311005;%%},
	title = {{Chemical freezeout and the QCD phase transition temperature}},
	volume = {B596},
	year = {2004}}

@article{deForcrand:2002ci,
	archiveprefix = {arXiv},
	author = {de Forcrand, Philippe and Philipsen, Owe},
	doi = {10.1016/S0550-3213(02)00626-0},
	eprint = {hep-lat/0205016},
	journal = {Nucl.Phys.},
	pages = {290-306},
	primaryclass = {hep-lat},
	reportnumber = {MIT-CTP-3270, CERN-TH-2002-102},
	slaccitation = {%%CITATION = HEP-LAT/0205016;%%},
	title = {{The QCD phase diagram for small densities from imaginary chemical potential}},
	volume = {B642},
	year = {2002}}

@article{DElia:2002gd,
	archiveprefix = {arXiv},
	author = {D'Elia, Massimo and Lombardo, Maria-Paola},
	doi = {10.1103/PhysRevD.67.014505},
	eprint = {hep-lat/0209146},
	journal = {Phys. Rev.},
	pages = {014505},
	primaryclass = {hep-lat},
	reportnumber = {GEF-TH-2002-12},
	slaccitation = {%%CITATION = HEP-LAT/0209146;%%},
	title = {{Finite density QCD via imaginary chemical potential}},
	volume = {D67},
	year = {2003}}

@article{MR1484478,
	author = {Bosma, Wieb and Cannon, John and Playoust, Catherine},
	doi = {10.1006/jsco.1996.0125},
	fjournal = {Journal of Symbolic Computation},
	issn = {0747-7171},
	journal = {J. Symbolic Comput.},
	mrclass = {68Q40},
	mrnumber = {MR1484478},
	note = {Computational algebra and number theory (London, 1993)},
	number = {3-4},
	pages = {235--265},
	title = {The {M}agma algebra system. {I}. {T}he user language},
	url = {http://dx.doi.org/10.1006/jsco.1996.0125},
	volume = {24},
	year = {1997}}

@article{Bzdak:2025rhp,
    author = "Bzdak, Adam and Koch, Volker and Vovchenko, Volodymyr",
    title = "{Acceptance dependence of factorial cumulants, long-range correlations, and the antiproton puzzle}",
    eprint = "2503.16405",
    archivePrefix = "arXiv",
    primaryClass = "nucl-th",
    doi = "10.1103/r6m1-b2tr",
    journal = "Phys. Rev. C",
    volume = "112",
    number = "2",
    pages = "024901",
    year = "2025"
}

\end{document}